\documentclass[aps,twocolumn,epsf,floats,pre,nofootinbib]{revtex4-1}
\usepackage{graphics,graphicx,epsfig}
\usepackage{amssymb,color}
\usepackage{epsf,epstopdf,wrapfig}
\usepackage{amsmath}
\usepackage{amsmath}
\usepackage{amssymb}
\usepackage{bbold}
\usepackage{graphicx}
\usepackage{bm}
\usepackage{xr}
\usepackage{scrextend}
\usepackage{verbatim}
\usepackage[usenames,dvipsnames,svgnames,table]{xcolor}
\usepackage{natbib}

\def\(({\left(}
\def\)){\right)}                       
\def\[[{\left[}
\def\]]{\right]}

\newcommand{\<}{\langle}
\renewcommand{\>}{\rangle}
\newcommand{\beq}{\begin{equation}}
\newcommand{\eeq}{\end{equation}}
\newcommand{\bea}{\begin{eqnarray}}
\newcommand{\eea}{\end{eqnarray}}

\begin{document}

\title{Propagating speed waves in flocks: a mathematical model}

\author{
Andrea Cavagna$^{1}$, 
Daniele Conti$^{1,2}$,
Irene  Giardina$^{1,2,3}$,
Tomas S. Grigera$^{4,5}$
}

\affiliation{$^1$ Istituto Sistemi Complessi, Consiglio Nazionale delle Ricerche, UOS Sapienza, 00185 Rome, Italy}
\affiliation{$^2$ Dipartimento di Fisica, Universit\`a\ Sapienza, 00185 Rome, Italy}
\affiliation{$^3$ INFN, Unit\`a di Roma 1, 00185 Rome, Italy}
\affiliation{$^4$ Instituto de F\'\i{}sica de L\'\i{}quidos y Sistemas Biol\'ogicos
CONICET -  Universidad Nacional de La Plata, 
  La Plata, Argentina}
\affiliation{$^5$ CCT CONICET La Plata, Consejo Nacional de Investigaciones Cient\'\i{}ficas y T\'ecnicas, Argentina}

\begin{abstract}
Efficient collective response to external perturbations is one of the most striking abilities of a biological system.
Signal propagation through the group is an important condition for the implementation of such a response.
Information transfer has been experimentally observed in the turning mechanism of birds flocks. 
In this context it is well-known also the existence of density waves: birds under predation, attempting to escape, give rise to self-organized density waves that propagates linearly on the flock.
Most aspects of this phenomenon are still not fully captured by theoretical models.
In this work we present a new model for the propagation of the speed (the modulus of the velocity) fluctuations inside a flock, which is the simplest way to reproduce the observed density waves. 
We have studied the full solution of the model in $d=1$ and we found that there is a line in the parameter space along which the system relaxes in the fastest way with no oscillation after a signal has passed. This is the critical damping condition.
By analyzing the parameters plane we show that critical damping represents an attractor for a steepest descent dynamics of the return time of the system.
Finally we propose a method to test the validity of the model through through future experiments.
\end{abstract}

\maketitle


\section{Introduction}

Many intriguing phenomena of the living world crucially depend on the interactions between the various components forming a biological system.
When the collective properties of a group emerge in an unpredictable way from the individual characteristics of the constituents one usually speaks of \textit{collective behavior} \cite{Ma_book,parisi_book,sethna2006statistical,sumpter_book,vicsek_review}. In physics, the emergence of collective behavior has been deeply investigated. Statistical mechanics proved to be a very powerful theory for understanding how macroscopic phenomena arise from the interaction of many microscopic components \cite{Ma_book,parisi_book,sethna2006statistical}. This success has raised the belief that it might be possible to use the same concepts and mathematical apparatus for describing the collective properties of biological systems. Indeed, thanks to the technological progress of recent years,  emergent biological phenomena are now susceptible of quantitative large-scale experiments and new challenges are opening up at the interface between physics and biology \cite{mora+al_11,bialek2015perspectives}.

An interesting feature of collective behavior in biological systems lies in the efficiency with which these systems are able to respond to stimuli coming from the external environment. This efficiency depends not only on the celerity with which the response is performed, but also on the time needed for the system to return to the stationary state, once the perturbation has passed. This is a crucial point, which has had little attention in the past: in order to transmit a signal across the group, each individual must be displaced from its original state (which is not necessarily a rest state); how the individual will go {\it back} to its original state? Clearly this is a relevant question, whose answer depends on finding a balance between transmitting the signal in the quickest way, but also disrupting the state of the system as little as possible.
This is the problem we investigate here.

The ability of group to respond as a whole has also important consequences on technological developments and control theory \cite {stear1987control,kube1992collective,jadbabaie2003coordination,leonard2007collective}.
Many biological systems at different scales display such behavior: bird flocks \cite{cavagna+al_10}, swarms of insects \cite{attanasi2014collective}, herds of mammals \cite{ginelli2015intermittent}, bacterial clusters \cite{dombrowski2004self,zhang2010collective}, cells \cite{szabo2006phase}, fish schools \cite{strandburg2013}, ant trails \cite{goss1989self}, etc., even at the human social level it is possible to find such features \cite{cont2000herd,helbing2001self}.
Although these systems have many characteristics in common, it has not been possible yet to construct a general theory in which to catalog them all. Each case has its own special features, however, the hope is that trying to develop simple mathematical models that reproduce the fundamental traits of these behaviors could represent a first step toward a universal theory \cite{giardina2008cobag}.

Among biological systems, flocks of birds have recently attracted much attention and have been studied both theoretically and experimentally \cite{reynolds1987flocks, vicsek_review, toner1998flocks, bialek+al_12, hemelrijk2012schools, cavagna_review}.
Theorists have produced elegant models of flocking  \cite{vicsek+al_95,toner1998flocks,ramaswamy_review,marchetti_review}.
The development of novel methods for recovering the three-dimensional positions and velocities of individual birds in large flocks of starlings have provided new quantitative data and renewed the interest in the field \cite{ballerini+al_08b}.
Previous studies have highlighted the importance of the transfer of information mechanism to achieve of an efficient collective response if flocks \cite{attanasi+al_14,cavagna+al_15,procaccini+al2011,hemelrijk2015wave}.
In order for the group to respond to a perturbation felt only by some individuals, it is necessary that information flows within the group \cite{tkavcik2016information}. Therefore a rapid and robust information transfer is essential for facilitating cohesion and ensuring a rapid reorganization of the group upon predator attacks \cite{attanasi+al_14}. Propagation phenomena can occur and have been observed experimentally in different degrees of freedom. In  \cite{attanasi+al_14}, using the full three-dimensional birds trajectories, it has been shown that in flocks during collective turns  (either spontaneous or elicited by the predator's arrival) a few individuals start turning and the change in direction of motion - a localized disturbance - propagates linearly through the whole group with a very large propagation speed. On the contrary, mutual distances and individual speeds remain approximately constant on the time scale where the global turn is concluded (typically a few seconds). This is a clear example of transfer of purely directional information. Another vivid manifestation of propagation phenomena is the occurrence of density waves. Video observations on large flocks of starlings under predatory attack indeed show the formation of waves in proximity of the arriving predator and their eventual propagation through the group (in absence of any collective turn) \cite{procaccini+al2011}.

The theoretical explanations of these phenomena are various. As shown in \cite{attanasi+al_14,cavagna+al_15}, the propagation of purely directional information during collective turns is due to the presence of a behavioral rotational inertia and second order terms in the dynamical evolution of the flight directions. These terms produce a linear propagation law with a speed depending on polarization, in quantitative agreement with experimental data.  This propagation is independent of density fluctuations, which are indeed not observed during turns.
The origin of density waves could be due to different mechanisms. Models of collective motion and polar active systems \cite{vicsek+al_95, vicsek_review} display anomalous density fluctuations \cite{chate+al_08b} due to the non trivial coupling between directional and positional degrees of freedom in these systems. Most of these models indeed consider the individual flight speed as fixed: what makes the local density fluctuate is that locally ordered regions tend to move together. In the ordered phase this coupling gives rise to non-trivial density waves on the very large scale, which have been studied  using a hydrodynamic approach \cite{toner+al_95, toner1998flocks, toner+al_98, toner_review, marchetti_review}. While these Hydrodynamic Density Waves (HDW) are certainly relevant for a variety of active systems, it is not clear whether they fully account for what is observed in natural flocks, for the following reasons: 
(i) HDW relate to the essence of the hydrodynamic approach: this approach considers the limit $L, t \rightarrow \infty$, while we know that natural systems are often far from these limits and exhibit important collective phenomena, such as collective turns, over medium scales.
(ii) HDW have an anisotropic propagation: waves that have significant speed in the reference frame of the flock propagate mainly in the direction orthogonal to the motion of the flock, while longitudinal modes are suppressed. Even though experimental observations are not clear on this issue, they seem to suggest that what matters most in how the wave propagates is the direction of the arriving predator rather than the flight direction of the flock itself \cite{procaccini+al2011}. 
(iii) Since the speed of each individual is fixed, in the hydrodynamic theory the density waves are derived from the fluctuations in the orientations of the system particles velocities and are in fact coupled to these \cite{toner+al_95, toner1998flocks, toner+al_98}.

There are other possible mechanisms that generate density waves besides HDW. As mentioned above, most flocking models of self-propelled particles assume that  individual speeds  are fixed. However, as highlighted for example in \cite{bialek+al_14,hemelrijk2015scale, peruani2007self}, we can consider fluctuations not only in the orientations of the velocity, but also in the speed, i.e. in the modulus of the individual velocity. Empirical data show that, in addition to the fluctuations in the direction, also the fluctuations in the speed are long-range, that is the correlation is scale-free \cite{cavagna+al_10}. To reproduce such correlations one needs to explicitly allow for speed variability in the individual equation of motion \cite{bialek+al_14, hemelrijk2015scale}. In this context, it is reasonable to hypothesize that density waves have a contribution coming not from the orientations of the velocity, but from speed. In particular this seems the main mechanism by which it is possible to generate density waves that propagate longitudinally.

In this work we present a model for the collective motion of birds in a flock that takes into account fluctuations in the individual speed of flight and admit the {\it linear}  propagation of such fluctuations through speed waves. To do this, we follow a similar theoretical path as that used to describe the fluctuations in flight orientations \cite{attanasi+al_14, cavagna+al_15}. We derive a dynamic equation for the speed that turns out to have an analogous structure as the telegraph equation, i.e. the equation that describe the propagation of electromagnetic waves in telegraph cables \cite{smirnov1964course, webster2016partial,magnusson2000transmission}. This equation has a special point in the space of parameters in which it assumes a simpler form, very similar to that of a pure wave equation.
We highlight this interesting aspect by analyzing the dispersion relation. 
Furthermore we relate this point to a generalized critical damping of the system: for this particular value of the parameters, the system minimizes the return time to the steady state, without oscillating, after a perturbation has passed through. This last feature addresses the question raised above:
minimizing the return time to the steady state improves the efficiency of the collective response. Furthermore, the absence of oscillations in the speed is certainly advantageous, since oscillations would cause an unnecessary waste of energy. We confirmed the significance of critical damping by studying the full solution of the equation in $d=1$. Finally we suggest a method by which one can experimentally verify the fundamental hypotheses of the model.

\section{A new equation of motion for the speed}
We look for a simple mathematical model, which contains the essential features of speed waves in birds flocks.
Since we want to describe a propagating phenomenon, we would like the model to reproduce a generalized wave equation for the individual speed $v_i = |\textbf{v}_i|$.
As often in collective behaviour, our starting point will be the Vicsek model.

\subsection{Vicsek model}
The Vicsek model (VM) \cite{vicsek+al_95} assumes that each particle tries to align its flight direction with those of neighbors while
moving with constant speed $|\textbf{v}_i|=v_0$. The  dynamical equations read (in three-dimensions)
\begin{align}
  \mathbf{v_i}(t+1) &= v_0 \mathcal{R}_\eta \Theta \left[{\bf v}_i+ \sum_j  n_{ij} \mathbf{v}_j(t) \right],  \label{Vicsek}\\
  \mathbf{r_i}(t+1) &= \mathbf{r}_i(t) +  \mathbf{v}_i(t+1),
  \label{vicsek}
\end{align}
where $\Theta({\bf x})={\bf x}/|{\bf x}|$ is the normalization operator and $R_\eta$ rotates its argument randomly within a spherical cone centered at it and spanning a solid angle $4\pi\eta$. The vectors $\textbf{v}_i$ and $\textbf{r}_i$ are, respectively, the velocity and the position of the bird $i$, and $v_0$ is the {\it constant} speed of the particles.
The matrix $n_{ij}$ is the connectivity matrix, which defines the neighborhood of interaction of $i$ (metric \cite{vicsek+al_95} or topological \cite{ballerini+al_08,ginelli+al_10}).
It is possible to interpret this equation as if each bird changes its velocity following a social force $\textbf{F}_i= J \sum_{j\ne i} n_{ij} {\bf v}_j$, where - for the sake of generality and for future convenience   - we introduced the parameter $J$ setting the scale of such force ($J=1$ in the original VM). The VM is defined for discrete time-steps, and describes a Markovian kind of dynamics. If we consider the limit of small time increments, it would lead (by appropriately rescaling the force strength $J$ and the noise amplitude) to a first order equation in time for the velocities (see - e.g. \cite{cavagna2014dynamical,cavagna+al_15}).

\subsection{Pseudo-Hamiltonian description}
We note that the social force in the VM can be written as $F_i=- \delta \mathcal{H}/\delta \boldsymbol{\phi}_i$, where
\begin{equation}
\mathcal{H} = - \frac{J}{2} \sum_{i,j} n_{ij} \, \textbf{v}_i(t) \cdot \textbf{v}_j(t)
\label{hamiltonian}
\end{equation}
It is tempting to interpret $\mathcal{H}$ as a Hamiltonian for a dissipative Langevin spin dynamics (where the spins are played by the flight directions of the moving individuals), but, due to the active nature of the system, we have to be careful \cite{zwanzig_book}. Active matter systems are out of equilibrium, the constituents absorb and dissipates energy, therefore detailed balance is not valid.
As a consequence, the stationary probability distribution is not given by the Boltzmann weight $P(\{\textbf{v}\}) \neq e^{-\beta \mathcal{H}\{\textbf{v}\}}$.\footnote{Nevertheless, there are evidences that in some cases it is still possible a statistical physics approach \cite{fodor2016far, loi2008effective, marconi2015towards}.}

The activity of a system of self-propelled individuals comes from the rearranging of the interaction network: animals move relative to each other, changing neighbors over time \cite{cavagna+al_13}. Indeed the connectivity matrix depends on time through the positions $\textbf{r}_i(t)$, which change with the velocities $\textbf{v}_i(t)$, so that $n_{ij}=n_{ij}(t)$.
This interdependence between velocity and position is responsible for many interesting properties of the VM such as the lowering of the critical dimension from $d=3$ to $d=2$ \cite{toner_review} and the presence of anomalous density fluctuations. The hydrodynamic theory of Toner and Tu \cite{toner+al_95,toner1998flocks} takes into account this aspect providing a continuous description of a coarse grained velocity field ${\bf v}({\bf x})$ where the movement of the network is fully included through the introduction of a coarse grained density field $\rho(\textbf{r})$. 
However, as previously discussed,  this description might not suitable for the phenomenon that we want to describe,
because we are interested in characterizing fast information propagation across finite-size systems, while the hydrodynamic treatment relies on asymptotically long times and very long distances.
The choice of the relevant time scale is indeed fundamental in determining the correct model:
when considering long times the rearrangement of the network cannot be overlooked, but for phenomena occurring on medium-short times,  the coupling between positional and orientational degrees of freedom might not be yet effective. For example, in collective turns the positional network remains approximately unaltered while directional information quickly propagates. In this case it is the presence of inertial second order derivatives in the dynamical equations (usually disregarded on long timescales) that gives rise to the propagation law \cite{attanasi+al_14,cavagna+al_15}.
Besides, experimental data show that the local rearrangement of the network in natural flocks happens on time scales much larger than the local updating time of the velocities \cite{cavagna+al_13,mora2016local}. Birds are in a state of local quasi-equilibrium where local directional quantities relax very quickly, as if the network were fixed\footnote{This is the reason why inference methods based on  static probability distributions give equivalent results to a full dynamical inference \cite{mora2016local}.}. These results indicate that natural flocks - at least the ones we are able to quantitatively observe - live in a regime where network rearrangements are {\it slow}, i.e. they are below the hydrodynamic regime.
In this paper we will therefore explore possible mechanisms giving rise to speed and density waves even in absence of network rearrangements.\footnote{Above some crossover scale (i.e. for very large flocks) the system would eventually enter the hydrodynamic regime. This crossover, which depends on the microscopic parameters of the system, has been investigated in \cite{cavagna2015silent}.
}

In this regime it is possible to consider the approximation where the positional network is fixed. In this case, $\mathcal{H}$ gains the role of a pseudo-Hamiltonian, an effective representation of the forces and constraints acting on the degrees of freedom of the system, which effectively determines the probability distribution. 
We can therefore proceed in the following way: we start from the pseudo-Hamiltonian Eq.~(\ref{hamiltonian}) and modify it to allow for fluctuating speeds. Then, in the spirit of the Vicsek model, we associate to this Hamiltonian a dynamical equation for the velocities.

\subsection{Speed as the fundamental degree of freedom}

Speed fluctuations are usually neglected, and the modulus of the velocity is assumed equal to some $v_0$, the average speed of the flock, which is fixed by the birds' physiology.  To study fluctuations around this average, the hard constraint must be abandoned, and an anchoring term must be added to the pseudo-Hamiltonian, as done in \cite{bialek+al_14} and \cite{hemelrijk2015scale}.  
The pseudo-Hamiltonian describing the system then becomes:
\begin{equation}
    \mathcal{H} = \frac{J}{4 v_0^2} \sum_{ij} n_{ij} |\textbf{v}_i - \textbf{v}_j|^2 + \frac{g}{2 v_0^2} \sum_i (v_i - v_0)^2.
 \label{pseudoham}
\end{equation}
where $v_i=|{\bf v}_i|$. The first term describes the tendency of the individual velocities to adjust both direction and modulus to their neighbors, while the second forces the speed towards the mean physiological value $v_0$, introducing a speed control constant $g$.  

In the highly polarized flocking phase, all the individuals move approximately in the same direction $\bf n$. For natural flocks, for example, the polarization $\Phi=1/N |\sum_i {\bf v}_i/|v_i|$ is very large (of order $0.9$) and the relative fluctuations both in flight directions and in speeds are very small \cite{cavagna+al_10}. We can write ${\bf v}_i=v_i {\bf s}_i$ (with $|{\bf s}_i|=1$) and express both the flight direction and the speed in terms of the fluctuations: ${\bf s}_i=s_i^L{\bf n} + \boldsymbol{\pi}_i$ and $v_i=v_0+u_i$. Expanding in the $\boldsymbol{\pi}_i$ and $u_i$ one can easily see that to leading order the pseudo-Hamiltonian (\ref{pseudoham}) splits into two terms,
\begin{align}
\mathcal{H}& = \mathcal{H}_{\mathrm{or}}(\{\boldsymbol{\pi}_i\}) +\mathcal{H}_{\mathrm{sp}}(\{u_i\})  \\
\mathcal{H}_{\mathrm{or}} &= \frac{J}{4 v_0^2} \sum_{ij} n_{ij} |\boldsymbol{\pi}_i - \boldsymbol{\pi}_j|^2  \\
\mathcal{H}_{\mathrm{sp}}& = \frac{J}{4 v_0^2} \sum_{ij} n_{ij} ( {u}_i - {u}_j)^2 + \frac{g}{2 v_0^2} \sum_i u_i^2
\label{ham-sp}
\end{align}
one involving only the orientations and another involving only the speed fluctuations.  This decoupling allows us to focus on the speed part only. From now on, we will therefore forget about the flight directions and focus on the speed fluctuations (see \cite{cavagna+al_15} for a description of the orientational dynamics).

\begin{figure}[t]
\centering
\includegraphics[scale=0.26]{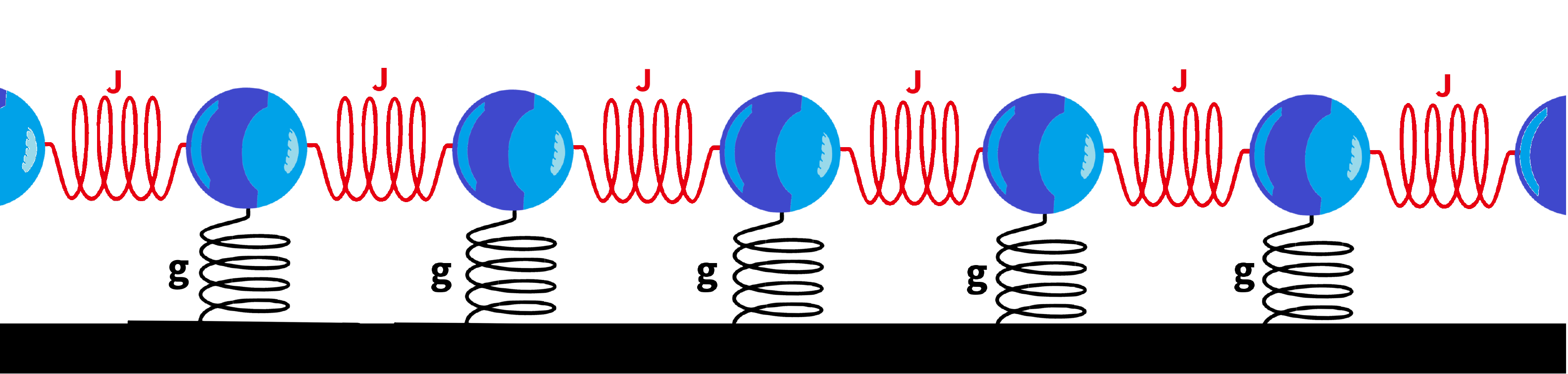}
\caption{Sketch of a chain of oscillator. Each oscillator, besides being connected to its first neighbors with strength $J$ (red spring), it is also tied to a base with strength $g$ (black spring), that forces it to have a determined position.}
\label{fig::oscillators}
\end{figure}

We notice that $\mathcal{H}_{\mathrm{sp}}$ is formally analogous to the Hamiltonian of a chain of harmonic oscillators in which every element, in addition to being connected to its nearest neighbors with strength $J$, has an additional spring that binds it to a fixed position,  \figurename\ref{fig::oscillators}. Of course, in this formal analogy, the degree of freedom $u_i$ is a displacement with respect to a certain reference {\it position}, while in our case $u_i$ is a displacement (or, more properly, a fluctuations) with respect to a certain reference {\it speed} (typically, the physiological speed of the individual). But despite the different intepretation of $u_i$, the analogy is exact and it provides a very useful paradigm we will refer to frequently in the rest of the paper.

Assuming that the variations in speed from bird to bird are smooth, we can take the continuous limit of this expression, in which the speed is a continuous function of the position in the flock, $\textbf{x}$, and the time. 
We can write $u(\textbf{x},t) =v(\textbf{x},t) - v_0$, and the Hamiltonian (\ref{ham-sp}) takes the form
\begin{equation}
    \mathcal{H}_{\mathrm{sp}} = \int \frac{d^3x}{a^3} \, \left \{ \frac{J a^2 n_c}{2 v_0^2} [\nabla u(\textbf{x},t)]^2 + \frac{g}{2 v_0^2} u^2(\textbf{x},t) \right \} \,.
    \label{ham-cl}
\end{equation}
The anchoring or control constant $g$ plays a fundamental role in determining the speed correlations. To see this we note that the pseudo Hamiltonian (\ref{ham-cl}) is Gaussian in the $u({\bf x})$. One can then easily compute the statistical equilibrium averages and get \cite{huang2009introduction}
\begin{equation}
    \langle u(\textbf{x})u(\textbf{x}')\rangle  \propto e^{-\frac{|\textbf{x}-\textbf{x}'|}{\xi}},
\end{equation}
where the correlation length is given by
\begin{equation}
    \xi \sim a \sqrt{J n_c/g}.
\end{equation}
In particular $\xi$ becomes infinite (and the correlation scale-free) at the critical point $g = 0$.  However, $g$ cannot be exactly zero, otherwise there is nothing to fix the mean speed of the birds.  Nevertheless for small enough values of $g$ the system is effectively critical due to the finite size effects $(\xi \propto L)$ \cite{bialek+al_14}.

A Langevin dynamics for speed fluctuations follows naturally from the pseudo-Hamiltonian just defined,
\begin{align}
    \eta \frac{\partial u(\textbf{x},t)}{\partial t} &= - \frac{\delta \mathcal{H}_{\rm sp}}{\delta u (\textbf{x},t)} + \zeta(\textbf{x},t) \notag \\
     &=  \frac{Ja^2 n_c} {v_0^2}\nabla^2 u(\textbf{x},t) -\frac{g}{v_0^2}u(\textbf{x},t) + \zeta(\textbf{x},t),
     \label{langevin}
\end{align}
where the friction $\eta$ is a constant that sets the time scale of the dynamics, and $\zeta(\textbf{x},t)$ is a random white noise $ \< \zeta(\textbf{x},t) \zeta(\textbf{x}',t) \>= 2\eta T a^3 \delta(\textbf{x} - \textbf{x}') \delta(t-t')$.  Approximating the network as fixed, this dynamics implies that  speeds follow a Boltzmann distribution $P(\{ u\}) \propto e^{-\beta \mathcal{H}_{\rm sp}(\{ u \})}$.  However, the structure of this equation of motion is very different from what we would expect from a propagating phenomenon.  Since $u$ is the fundamental degree of freedom, this equation is an overdamped first-order equation of the parabolic type \cite{strauss1992partial,webster2016partial}.  This means that information travels sublinearly, $x\sim \sqrt{D t}$, and that a speed of propagation cannot even be defined.  The diffusive structure of this equation is therefore unsuitable to describe the propagating phenomenon we expect.

\subsection{Symmetry generator and inertia}

To obtain a new theory able to describe propagating speed waves we switch to an underdamped Hamiltonian dynamics.  
In the previous section we considered an overdamped Langevin dynamics for the speed. However, in the limit of zero noise and dissipation, one would like the speeds to obey some deterministic dynamics ruled by the forces at play in the system (i.e. mutual alignment and anchoring). The most obvious possibility is a Hamiltonian dynamics, which has the advantage of automatically implement the symmetries present in the system, and it has proven to be the key ingredient to reproduce propagation waves in the orientational degrees of freedom \cite{attanasi+al_14,cavagna+al_15}. To this aim, we introduce a canonical pair of coordinates $(u, w)$, where $u$ are the speed fluctuations and $w$ is the generator of the transformation parametrized by $u$. It is defined by the Poisson relation
\begin{equation}
    \frac{df}{du}= \{f, w\}.
\end{equation}
This equation states that the variation of any observable $f$ under the transformation parametrized by $u$ is given by the Poisson bracket of $f$ with the generator $w$.  This transformation corresponds to a translation in the speed and it is the fundamental mechanism generating speed waves.
Once introduced the conjugated momentum $w$, we can build the full Hamiltonian for $u$ and $w$ by adding to the interaction term containing the speeds (${\mathcal H}_{\rm sp}$), a generalized kinetic term,
\begin{equation}\begin{split}
    \mathcal{H} =  \int \frac{d^3x}{a^3} \,\biggl\{  &\frac{J a^2 n_c}{2 v_0^2} [\nabla u(\textbf{x},t)]^2 + \frac{g}{2 v_0} u^2(\textbf{x},t) + \\    
 &\, \frac{w^2(\textbf{x},t)}{2 \mu}\biggr\},
\end{split}\end{equation}
where $\mu$ is the inertia associated to the canonical pair $(u,w)$.
It is important to note that $\mu$ is not the standard mass, but a generalized inertia that embodies the resistance of the bird to a change of $\dot{v}$.
A reference to our chain of oscillators may be of help here: if we interpret $u$ as a displacement, then $w$ is simply the regular momentum, which generates the space translations parametrized by $u$, and $\mu$ would be the normal mechanical mass. Note, in this context, that the term $gu^2$ breaks the translational symmetry, because each particle has a preferred position thanks to it. Now let us switch to the interpretation in which $u$ is a speed fluctuation, rather than a space displacement. In this case the symmetry generated by $w$ and parametrized by $u$ is still a translation, but a translation in the space of {\it speed}, which we may call a {\it boost}. The term $gu^2$ breaks this symmetry, thus giving to each individual a preferred speed, its physiological value.
The interesting point is that, when a system is highly polarized, the boost transformation we are talking about (i.e. a shift in the speed) becomes conceptually quite close to a Galilean transformation (clearly, this is far from true if the polarization is low: a constant shift of each speed does {\it not} produce a uniform velocity shift). Because flocks are highly polarized, this analogy is fair and in this context we see then that the $gu^2$ term breaks Galilean invariance: the equations of motion are not the same in any inertial reference frame, because birds have a physiological reference speed. Indeed birds move through a resistive medium, which provides a special Galilean reference frame, where the dynamics is simpler and different from those in other reference frames \cite{toner_review}. This has the rather interesting consequence that the scale-free point $g=0$ identifies with the point which restores Galilean invariance in the system.

\subsection{Speed waves}
Having an inertial term allows us to consider a dynamics, given by the canonical equations of motion:
\begin{align}
    \frac{\partial u(\textbf{x},t)}{\partial t} &= \frac{\delta \mathcal{H}}{\delta w(\textbf{x},t)} \label{ham1}\\
    \frac{\partial w(\textbf{x},t)}{\partial t} &= - \frac{\delta \mathcal{H}}{\delta u(\textbf{x},t)} \label{ham2}
\end{align}
We can now reinstate friction and noise, to get a set of equations containing both conservative and dissipative terms:
\begin{align}
    \frac{\partial u(\textbf{x},t)}{\partial t} &= \frac{\delta \mathcal{H}}{\delta w(\textbf{x},t)}\\
    \frac{\partial w(\textbf{x},t)}{\partial t} &= - \frac{\delta \mathcal{H}}{\delta u(\textbf{x},t)} - \eta \frac{\partial u(\textbf{x},t)}{\partial t} + \zeta(\textbf{x},t).
\end{align}
\begin{figure*}[t!]
\centering
\includegraphics[width=0.45\textwidth]{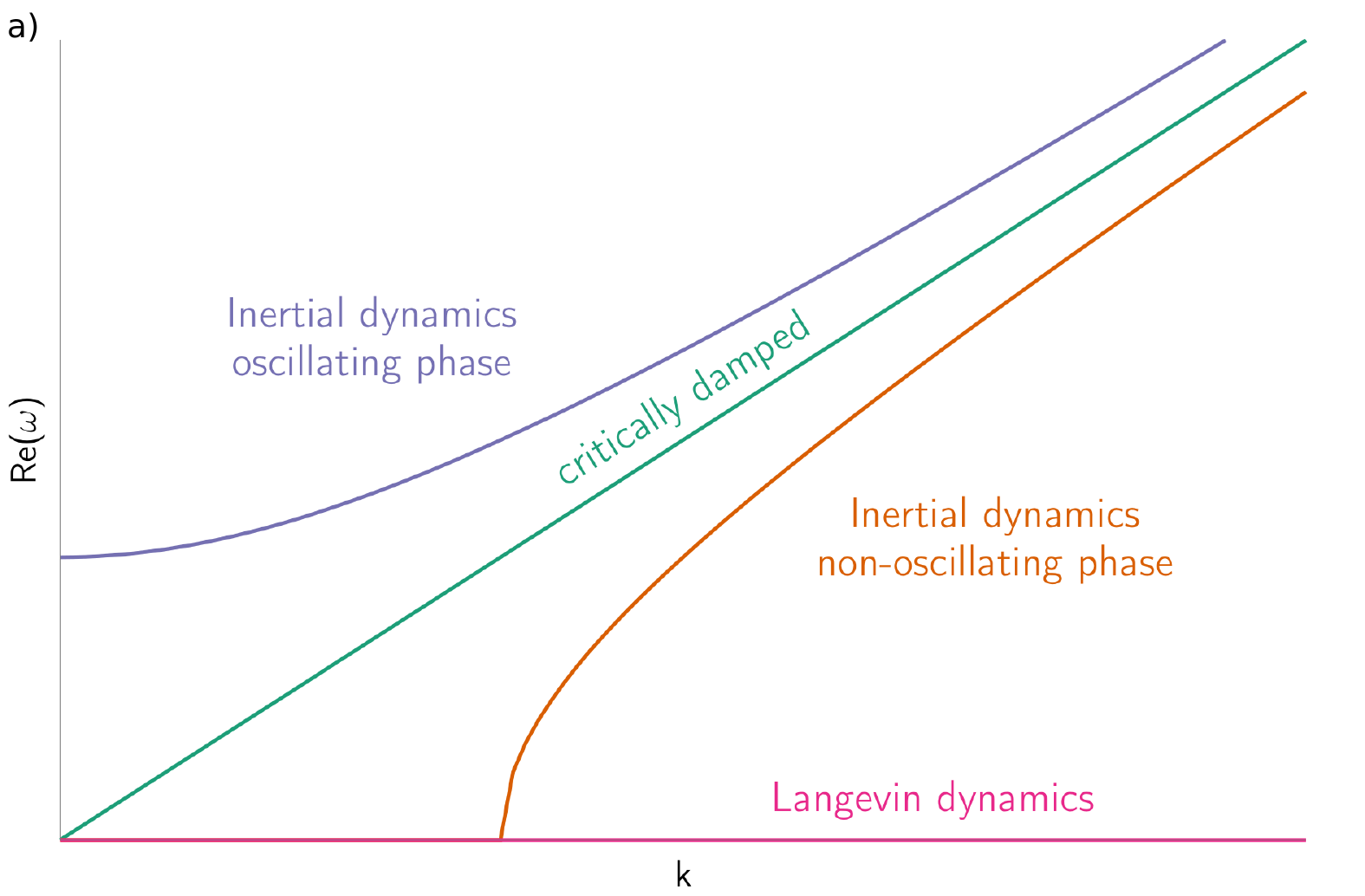}
\quad\quad
\includegraphics[width=0.45\textwidth]{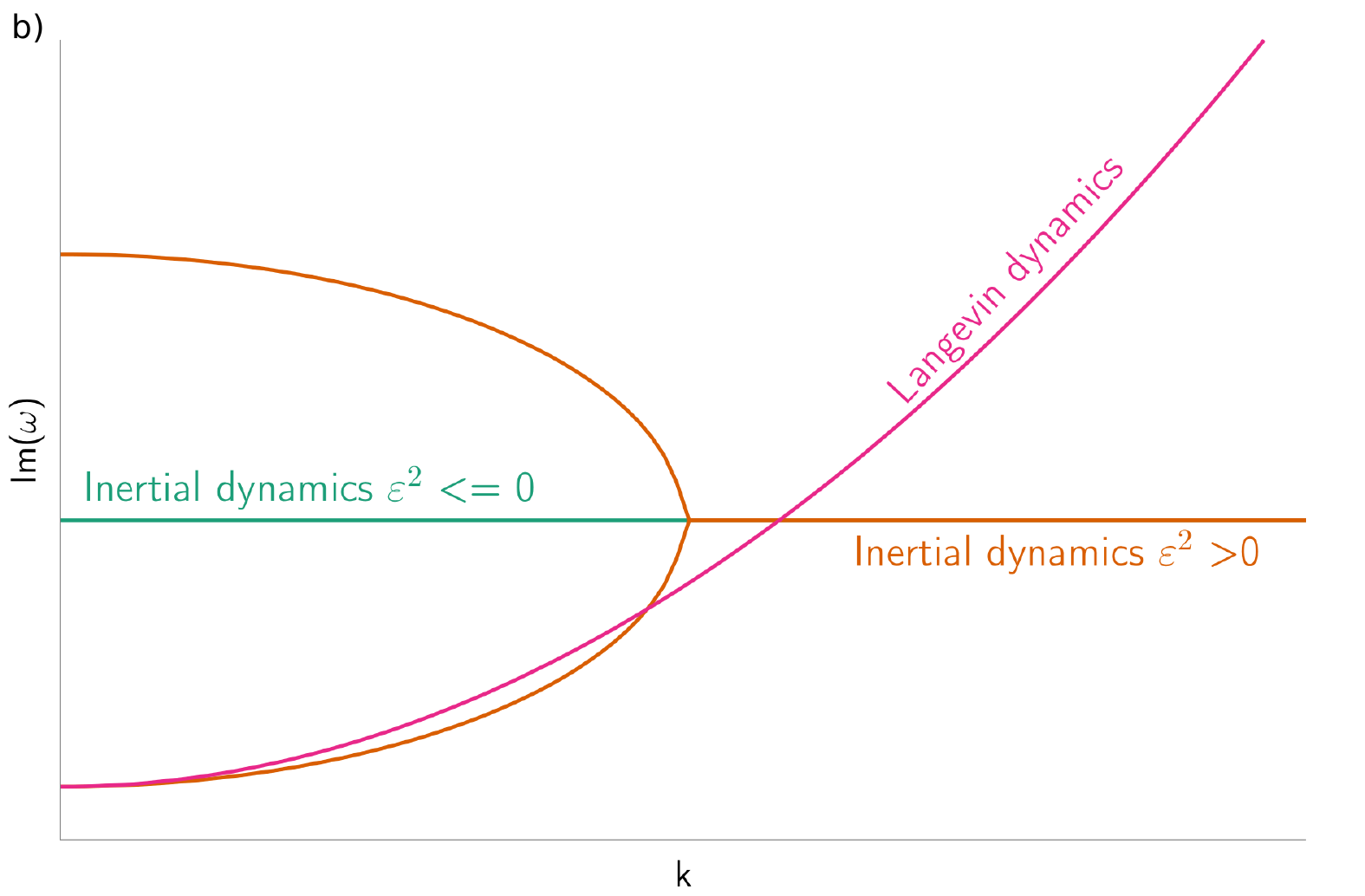}
\caption{Sketch of the dispersion law. (a) It is drawn the real part of the frequency $\omega$ for different cases. If $\varepsilon^2 < 0$ (lilac line) we are in the oscillating phase and there is propagation for every value of $k$; if $\varepsilon^2 >0$ (orange line) the system is non-oscillating and there is propagation only for $k>k_0= \frac{\varepsilon}{c}$; if $\varepsilon^2 = 0$ (teal line) we are at the critical damping and there is always linear propagation; for Langevin dynamics the real part of the frequency is zero. (b) It is drawn the imaginary part of the frequency $\omega$ for different cases. The teal line represent the oscillating and critically damped situations in which $\mathrm{Im}(\omega)$ is constant; the orange line represent the non-oscillating regime in which the $\mathrm{Im}(\omega$) is constant only for $k > k_0$ and grows quadratically with $k$ for small values of $k$. }\label{fig:dispersion}
\end{figure*}From this pair of equations follows the equation of motion for the speed,
\begin{equation}
\mu \frac{\partial^2 u(\textbf{x},t)}{\partial t^2} + \eta \frac{\partial u(\textbf{x},t)}{\partial t} + gu(\textbf{x},t) - a^2 J n_c \nabla^2 u(\textbf{x},t) = \zeta(\textbf{x},t)
\end{equation}
which is a second order equation of the hyperbolic type, suitable to represent propagating phenomena \cite{strauss1992partial}.
A better insight of this equation can be gained by rewriting it as
\begin{equation}
\frac{\partial^2 u(\textbf{x},t)}{\partial t^2} + 2 \gamma \frac{\partial u(\textbf{x},t)}{\partial t} + \omega_0^2 u(\textbf{x},t) - c^2 \nabla^2 u(\textbf{x},t) = \zeta (\textbf{x},t),
\label{full}
\end{equation}
where $c^2= J a^2  n_c/\mu$ is the phase velocity of the propagating waves, $\gamma = \eta/2\mu$ is the reduced friction and $\omega_0^2=g/\mu$ is the natural frequency (i.e.\ the frequency with which the system would oscillate in the absence of the social force).  This equation is known in the literature as the telegraph equation  \cite{smirnov1964course, webster2016partial}.  It can be further simplified by introducing a new field $\varphi(\textbf{x},t) = e^{-\gamma t}u(\textbf{x},t)$ in such a way that the terms containing $\partial  \varphi / \partial t$ drop out in the equation for $\varphi$.
Then for the homogeneous case we get
\begin{equation}
\frac{\partial^2 \varphi(\textbf{x},t)}{\partial t^2} = c^2 \nabla^2 \varphi(\textbf{x},t) + \varepsilon^2 \varphi(\textbf{x},t), \label{eq:1}
\end{equation}
where,
\beq
\varepsilon^2= \gamma^2 - \omega_0^2 \ .
\eeq
Note that $\varepsilon^2$ can be positive or negative depending on whether the friction dominates over speed control or vice-versa; the use of the square notation derives from the definition given in the standard telegraphic equation, where $\varepsilon^2$ is defined semipositive.  
In this form it is clear that $\varepsilon^2=0$ is a critical value: if the parameters are such that $\varepsilon^2 = \eta^2/4 \mu^2 - g/\mu= 0$, Eq.~\eqref{eq:1} reduces to the classical wave equation.
This case in which the physical constants can be adjusted to eliminate the dispersion corresponds in literature to the loss-less transmission line case \cite{magnusson2000transmission}.
Besides reducing the distortion of the signal, this point has the interesting property of minimizing the time required for the system to return to the unperturbed state.
It thus represents an optimal situation for the information transfer.
From a biological point of view it would be an extremely useful mechanism.
It is therefore important to examine in depth the working principle and properties of this mechanism.

\section{The dispersion relation}

In order to better understand the differences between the Langevin dynamics (Eq.~(\ref{langevin}) and the Hamilton dynamics (Eqs.~(\ref{ham1},\ref{ham2})) and their role in the complete dynamical equation (\ref{full}), it is useful to study the dispersion relation.
This can be obtained by solving the differential equations with the Green's function method in Fourier space \cite{morse1946methods}.
Given a differential equation
\begin{equation}
    \mathcal{L}(\textbf{x},t) u(\textbf{x},t) = \zeta (\textbf{x},t),
\end{equation}
where $\mathcal{L}(\textbf{x},t)$ is a generic differential operator, the Green's function $G(\textbf{x},t)$ is defined as the function such that
\begin{align}
    \mathcal{L}(\textbf{x},t) G(\textbf{x},t) &= \delta (\textbf{x},t)  , \\
    u(\textbf{x},t) &= \int d \textbf{x}' dt' G(\textbf{x} - \textbf{x}',t - t') \zeta (\textbf{x}',t')  .
\end{align}
If $\mathcal{L}$ is linear, its Fourier space counterpart becomes a polynomial in the frequencies $\omega$ and momenta $k$, and the Green's function is simply the reciprocal of this polynomial. The poles of the Green's function give the relationship between frequency and momentum that must be fulfilled by the (possibly damped) plane waves that can propagate in the system: this is the dispersion relation.  

\subsection{Langevin vs Hamilton dynamics}

For Langevin dynamics the dynamic equation is of first order in the time, and in consequence the frequency is purely imaginary:
\begin{equation}
\omega = i (D k^2 + \omega_0) \ , 
\label{zuppa}
\end{equation}
where, 
\beq
D=(J a^2 n_c)/(v_0 ^2 \eta) \quad , \quad \omega_0 =g/(v_0 \eta)
 \ .
\label{mazza}
\eeq
The vanishing of the real part corresponds to the fact that there is no propagation, but only exponential damping.  The (imaginary) frequency has a gap  $\omega_0$  plus a quadratic diffusive term $Dk^2$.
As a consequence all the modes are overdamped and a disturbance spreads diffusively through the system. 

Introducing the conjugate momentum of the speed and the generalized inertia gives, as we saw, an equation of second order in time and space, eq.\eqref{eq:1}.  The dispersion polynomial associated to it is of second order both in frequency and momentum, 
\begin{equation}
\omega = i \gamma \pm \sqrt{c^2 k^2 - \varepsilon^2} \, .
\label{togotogo}
\end{equation}
From this dispersion law it is clear that the parameter $\varepsilon^2$ plays a fundamental role in determining the type of propagation.
If $\varepsilon^2 < 0$ the argument of the square root is always positive, and $\omega$ has a real part even for $k=0$, $\mathrm{Re}\,\omega(k=0)= \pm |\varepsilon|$.
This is the oscillating zone: there is propagation for every $k$, and the dispersion relation is quadratic for small $k$, approaching a linear behavior at large $k$. 
On the other hand if $\varepsilon^2 >0$, the argument of the square root changes sign with $k$.
In this case the system is non-oscillating and there is propagation only for  $k>k_0= \varepsilon/c$.
However for large values of $k$ we recover again a linear dispersion law.

It is crucial to note that the particular value $\varepsilon^2 = 0$ guarantees linear propagation at {\it all} values of $k$, $\mathrm{Re}\, \omega= \pm ck$, but with some damping $\gamma$, independent of $k$.
In this case the real parts of the two roots coincide since the damping factor $\gamma$ and the natural frequency $\omega_0$ of the system perfectly balance.
We have already noticed that the speed Hamiltonian is analogous to the one of a chain of oscillators. The picture we have just described - for $\varepsilon^2=0$ - has an interesting connection with what happens even in a {\it single} damped harmonic oscillator, and in particular with the definition of \emph{critical damping}.
 It is useful to explore the meaning of this toy case in order to simply catch the fundamental properties of this particular value.

\subsection{Toy model: critical damping and minimum return time}
The damped harmonic oscillator (DHO) well represents many different physical situations (mechanical oscillator, LRC circuit, etc.).
In this case we would like to use it as a paradigmatic situation for what happens in the speed waves model we derived.
The well-known equation of motion is \cite{taylor2005classical}:
\begin{equation}
 m \ddot{x}(t) = -\eta \dot{x}(t) -kx(t) \, ,
\end{equation}
where $x(t)$ is a generalized coordinate function of time, $m$ is the inertia, $\eta$ the viscosity and $k$ the elastic constant, or stiffness.
In order to make the comparison with the speed waves model clearer, we can rewrite in the following way:
\begin{equation}
 \ddot{x}(t) +  2\gamma \dot{x}(t) + \omega_0^2 x(t) =0\, .
\end{equation}
where we have introduced the damping constant $\gamma= \eta/2m$ and the natural frequency $\omega_0=\sqrt{k/m}$.
The equation of the DHO does not refer to a field, but to a single coordinate, and lacks the propagating term in $k$.
However in this context we are not interested in these aspects; what
we want to understand is how the different relationship between $\gamma$ and $\omega_0$ (and hence the value of $\varepsilon^2$)
determines the way the system returns to (mechanical) equilibrium. Using again the Green's function method we can obtain the dispersion polynomial,
\begin{equation}
\omega = i \gamma \pm \sqrt{\omega_0^2 - \gamma^2} = i \gamma \pm \sqrt{-\varepsilon^2} \, .
\end{equation}
The shape of the solution depends crucially on the value of $\varepsilon^2$, that is on the balance between reduced viscosity $\gamma$ and natural frequency $\omega_0$.
There are two different solutions separated by a critical point.
For $\varepsilon^2 <0$ we are in the underdamped regime, meaning that inertia (and stiffness) dominate over viscosity; since the real part is large, here the solution displays a clear oscillatory behaviour.
For $\varepsilon^2 >0$ the DHO enters in the overdamped regime, where the two roots are purely imaginary.
In this regime viscosity dominates  and the solution does not show oscillations, but falls to zero exponentially.

At precisely $\gamma = \omega_0$, namely $\varepsilon^2 =0$ one has \emph{critical damping}, which represents the boundary between underdamping and overdamping. As in the overdamped case, the solution shows no oscillations, but the peculiarity of this particular condition is that the system relaxes a perturbation as quickly as possible, minimizing the return time $\tau$ to the rest position.
A critically damped system therefore relaxes a perturbation as fast as possible, without oscillating \cite{taylor2005classical}.
Of course, our full equation for the speed is more complicated than this simple case because we have a field (infinite degrees of freedom), rather than one degree of freedom; this is why the extra term $k^2$ arises in the full dispersion relation \eqref{togotogo}. Yet critical damping as displayed by a single oscillator is a very useful intuitive concept also for the more complicated case.

There are many situations in which one wants passing disturbances to end as quickly as possible (shock absorbers of a car, closing system of a door, etc.).
In all these cases it is necessary to adjust the parameters so that the damping is as close as possible to critical.
Often the mechanisms we observe in nature are in a minimum state of a mathematical function used to describe the physical problem.
In particular in the case of a collective response it is very important that the reaction to external perturbations is performed in the shortest possible time, spending as little energy as possible.
In particular, such an optimization seems sensible in the case of a flock in motion: it would appear reasonable to avoid situations where, after responding to a perturbation, a particular bird would start oscillating around the cruising speed, or take a very long time to return to that value.  A critical damping on the propagation of speed fluctuations would ensure a cohesive and efficient movement.
We will now investigate how this intuition is supported by the solution of the speed waves model.

\section{Exact solution of the speed wave equation in \boldsymbol{$d=1$}}
We now study the full solution of the speed waves model.
We start from the simplest case, that is, from the solution in dimension $d=1$.
Again for reasons of simplicity we will assume that the system is infinite.
Although this may seem an unrealistic approximation, it has no consequences for the purpose of the study, since our interest here is the way the signal propagates through space, and this propagation does not rely on the infinite nature of the system.
In particular, we would like to understand if and how a critical damping regime is reflected by the mathematics of the problem.

\subsection{Wave and wake}
\begin{figure}[htbp]
\centering
\includegraphics[width=0.45\textwidth]{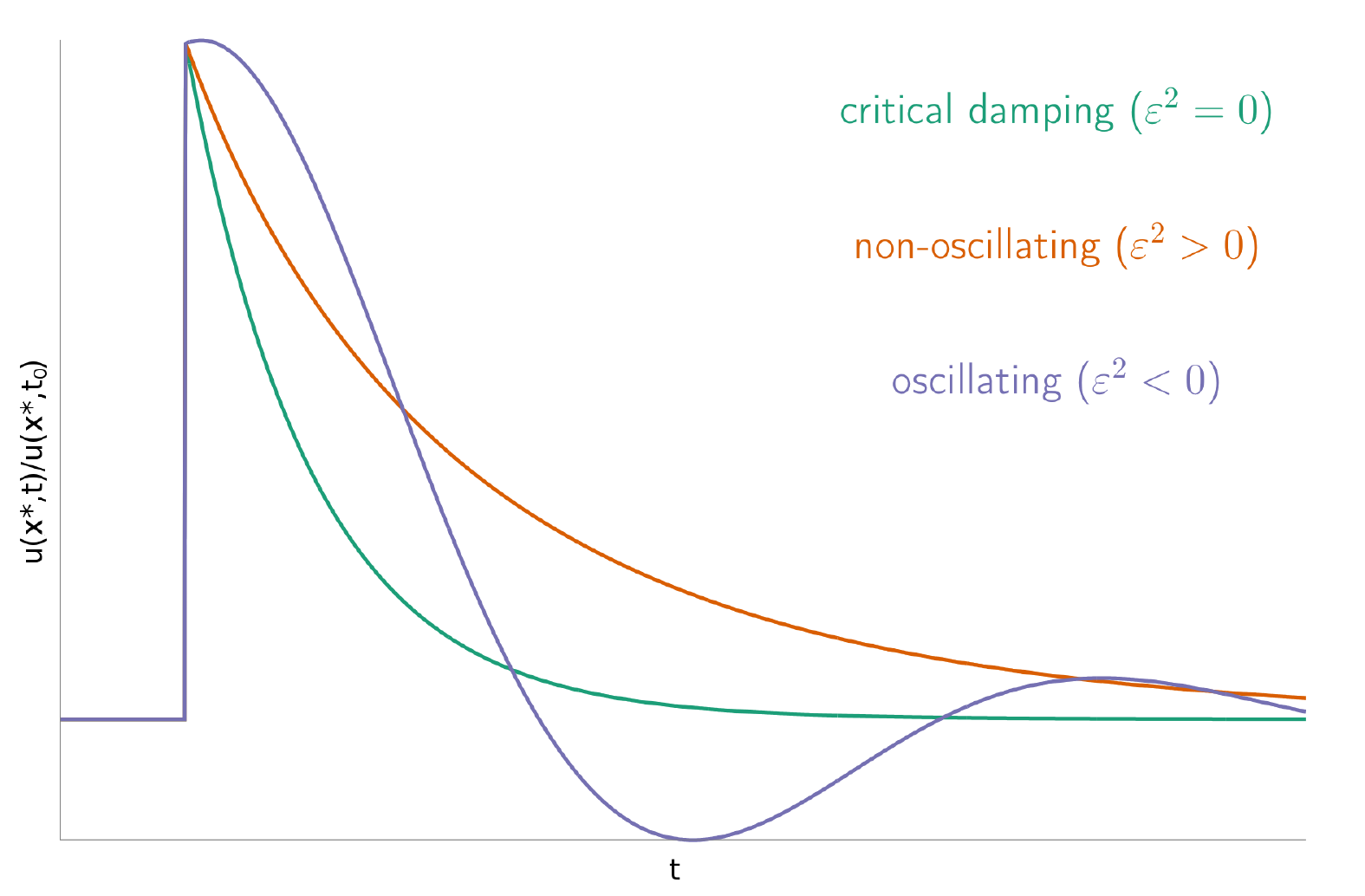}
\caption{Different solution of the speed equation in $d=1$. Depending on the value of $\varepsilon^2$ the solution will go to zero differently: at the critical damping point $\varepsilon^2=0$ (teal line) it reaches the zero in the fastest way without oscillating; in the non-oscillating regime (orange line), $\varepsilon^2 >0$ the solution goes to zero more slowly while in the oscillating case (lilac line) the field displays oscillations before going to the original value.}
\label{fig::damping}
\end{figure}

The solution for general initial conditions,
\begin{equation}
 u(x,t)|_{t=0}= \phi(x), \qquad \left. \frac{\partial u(x,t)}{\partial t} \right |_{t=0}= \psi(x),
\end{equation}
is given by \cite{smirnov1964course,webster2016partial},
\begin{equation}\begin{split}    
    &u(x,t) =e^{-\gamma t} \Biggl\{ \frac{\phi(x-ct) + \phi(x+ct)}{2} \\
    &+ \frac{1}{c}\int^{x+ct}_{x-ct} [\gamma \phi(x') + \psi(x')] \mathrm{I}_0\left(\frac{\varepsilon}{c}\sqrt{c^2t^2 - (x'-x)^2} \right) dx' \\
    &- \frac{\varepsilon t}{2}\int^{x+ct}_{x-ct} \frac{\phi(x') \, \mathrm{I}_1\left(\frac{\varepsilon}{c}\sqrt{c^2t^2 - (x'-x)^2} \right)}{\sqrt{c^2t^2 - (x'-x)^2}}  dx'\Biggr\}    
\end{split}\end{equation}
where $\mathrm{I_0}$ and $\mathrm{I_1}$ are modified Bessel function of the first kind.
Since we are interested in the way a localized perturbation propagates, we consider the initial conditions with a pulse at $x=0$:
\begin{equation}
 u(x,t)|_{t=0}= u_0\delta(x), \qquad \left.\frac{\partial u(x,t)}{\partial t}\right|_{t=0} = 0.
\end{equation}
In this case we can write the solution as
\begin{equation}
\begin{split}    
    u(x,t) = u_0 e^{-\gamma t} &\Biggl\{ \frac{\delta(x-ct) + \delta(x+ct)}{2} + \\
    &+ \biggl[\frac{\gamma}{c} \mathrm{I}_0 \left(\frac{\varepsilon}{c}\sqrt{c^2t^2 - x^2} \right) + \\
    &- \frac{\varepsilon t}{2}\frac{\mathrm{I}_1\left(\frac{\varepsilon}{c}\sqrt{c^2t^2 - x^2} \right)}{\sqrt{c^2t^2 -x^2}}  \biggr] \theta(|x| - ct) \Biggr\}  .
    \end{split}
\label{tel_I}
\end{equation}
For $\varepsilon^2 <0$ the modified Bessel functions can be replaced by Bessel functions of the first kind $\mathrm{J_0}$ and $\mathrm{J_1}$:
\begin{equation}
\begin{split}    
    u(x,t) = u_0 e^{-\gamma t} &\Biggl\{ \frac{\delta(x-ct) + \delta(x+ct)}{2} + \\
    &+ \biggl[\frac{\gamma}{c} \mathrm{J}_0 \left(\frac{\varepsilon}{c}\sqrt{c^2t^2 - x^2} \right) + \\
    &+ \frac{\varepsilon t}{2}\frac{\mathrm{J}_1\left(\frac{\varepsilon}{c}\sqrt{c^2t^2 - x^2} \right)}{\sqrt{c^2t^2 -x^2}}  \biggr] \theta(|x| - ct) \Biggr\}.
    \end{split}
\label{tel_J}
\end{equation}
Finally, for $\varepsilon^2=0$ the solution reduces to
\begin{equation}
 u(x,t)= u_0 e^{-\gamma t} \left\{ \frac{\delta(x-ct) + \delta(x+ct)}{2} + \frac{\gamma}{c}\theta(|x| - ct) \right\}. \label{eq:3}
\end{equation}
The main effect of viscosity is the presence of the overall damping factor $e^{-\gamma t}$.  Looking at the terms within braces, we see that the first two terms represent pulses propagating left and right with speed $c$; this 
term would be present also without anchoring (standard wave equation).  The remaining term instead introduces a new phenomenon: the wave leaves a wake.  Even after the wave front has passed, an effect that originates from all the points where the initial condition is different from zero is present at all points within a distance $t/c$ from them.  This wake vanishes exponentially in time \cite{smirnov1964course,webster2016partial}.  As a consequence a given point does not return instantaneously to its equilibrium position (as it would in a d'Alembert wave), but there is a tail in time, the structure of which depends on the value of the parameters (see \figurename \ref{fig::damping}).

\subsection{Return time}

\begin{figure*}[t!]
\centering
\includegraphics[width=.9\textwidth]{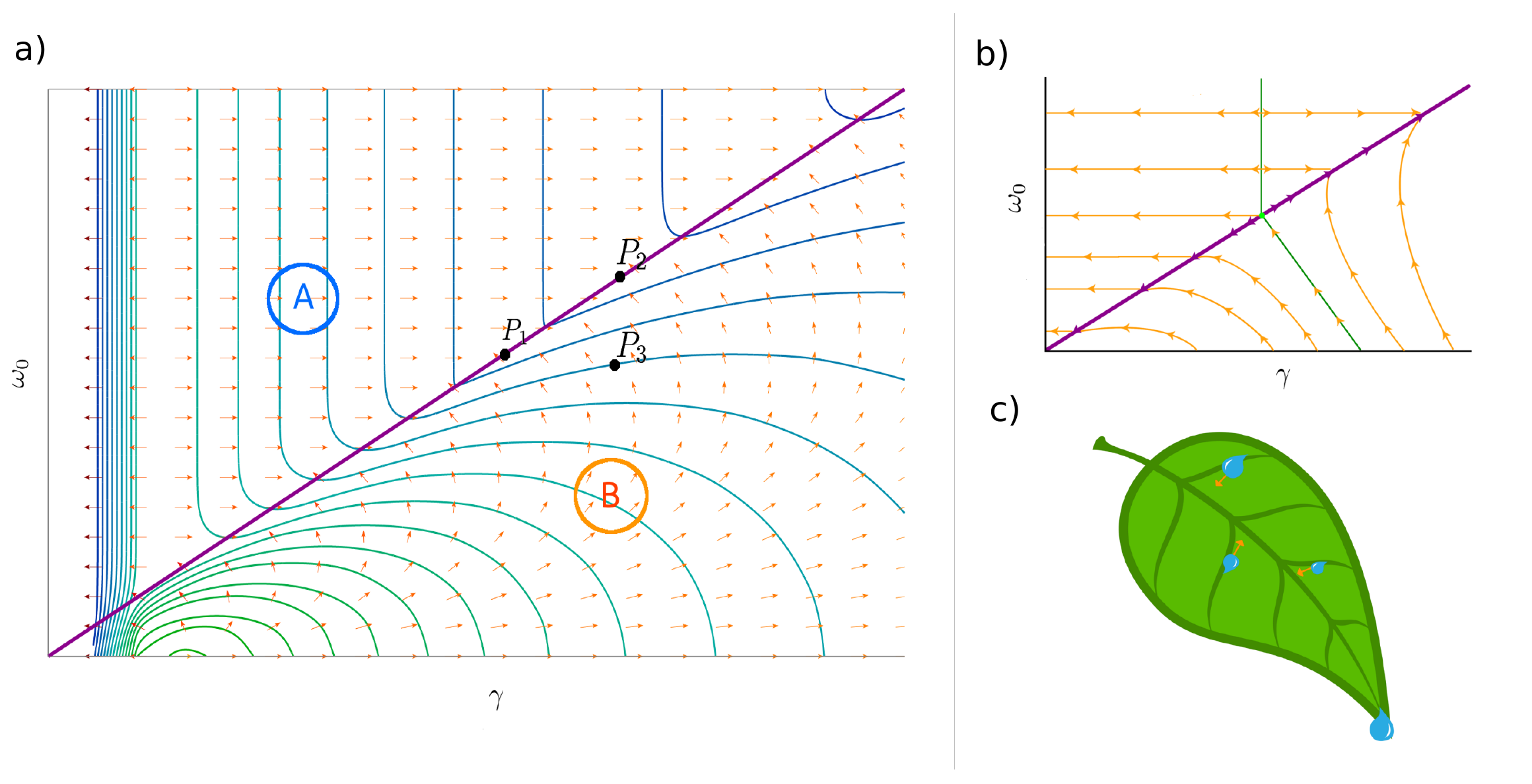}
\caption{(a) Level lines of $\tau(\gamma, \omega_0)$ are drawn with colors from blue for lower values, to green for the higher ones. In the same figure there is the vector field given by the negative gradient of the function, also here the color is the intensity of the field, ranging from yellow for lower values to dark red for the highest ones. The violet line represents the critical line. (b) Zoom of the gradient flow near the saddle point dividing the transparent zone from the intermediate zone. The green line represents the separatrix which divide the steepest descent dynamics. (c) A sketch of the dynamics of a water drop on a leaf: water drops tend to go toward the leaf rib.}
\label{fig:gradient}
\end{figure*}

To understand how the field returns to its unperturbed value, it is necessary to define the return time $\tau$.  In the presence of noise, equilibrium fluctuations will have an amplitude
\begin{equation}
\lim_{t \to \infty} u(x,t) = \sqrt{\langle u^2(x,t) \rangle} = \sqrt{T/g}  ,
\end{equation}
where $T$ is the effective temperature.  Hence we will define $\tau$ as the time it takes the for the solution to decay to certain level $1/q$ (proportional  to the level of noise fluctuations) after the arrival of the wavefront.  Since the signal arrives at $x^\star$ at a position-dependent time $t_0=x^{\star}/c$, our definition for the return time at $x=x^\star$ reads
\begin{equation}
  \label{eq:2}
   u(x^\star,t_0(x^\star)+\tau) = 1/q.
\end{equation}
This definition is appropriate for the non-oscillating phase, but when $\varepsilon^2 <0$ the solution oscillates and has an infinity of zeros and the above definition would not give a unique value of $\tau$; on the other hand, taking the smallest solution is not appropriate because it will be dominated by the period of the oscillations (at high frequency at least).  Therefore in the oscillating case we simply disregard oscillations and assume that the return time is determined by the exponential decay envelope.
Hence for $\varepsilon^2<0$ we ignore the oscillating Bessel functions and find
\begin{equation}
 \tau = - t_0 +\frac{1}{\gamma}\log\left(\frac{u_0q\gamma}{c}\right) .
\label{tauzero}
\end{equation}
This solution is also good for the critical line $\varepsilon^2=0$ as can be seen from Eq.~\eqref{eq:3}.

Along the critical line, the return time $\tau$ is characterized by a first region where it is zero, followed by a zone in which it grows up to a maximum at $\gamma= c e/u_0 q$, and then decreases until it vanishes again.  However, the two regions of vanishing $\tau$ are qualitatively different.
The first region, which we call the \emph{transparent zone}, is characterized by very small values of $\gamma$.
The height of the wake is proportional to $\gamma$, so all the wake falls below the noise fluctuation level, and the only relevant perturbation is the traveling $\delta(x \pm ct)$, which has $\tau =0$; we conclude that in the transparent region, the propagation is effectively d'Alembert.
The second region where the return time is zero is instead characterized by very large values of the damping $\gamma$.  This we call the \textit{opaque zone}: here $t_0 \gg\ 1/\gamma$, therefore the signal is strongly damped and cannot reach the position $x^{\star}$, since the amplitude of all the terms is below the noise threshold. 

We do not have an analytic expression for $\tau$ for $\varepsilon^2>0$, but close to the critical line we can expand the Bessel functions for small values of the argument and solve Eq.~\eqref{eq:2} recursively to get
\begin{equation}\begin{split}
 \tau =& \,  \tau_0 +\frac{\varepsilon^2}{2\gamma^2}\left(\frac{t_0}{2} + \frac{\tau_0}{2} +\gamma t_0\tau_0 + \gamma \frac{\tau_0^2}{2}\right) \\
=&\, t_0 +\frac{1}{\gamma}\log\left(\frac{u_0q\gamma}{c}\right) \\
 &+ \frac{\varepsilon^2}{4 \gamma^3}\left\{\log\left(\frac{u_0q\gamma}{c}\right)\left[1 + \log\left(\frac{u_0q\gamma}{c}\right)\right] - \gamma^2 t_0^2\right\} \, .
\label{taue}
\end{split}\end{equation}
This expression shows that $\tau$ grows when $\varepsilon^2$ grows at fixed $\gamma$.  Hence, $\varepsilon^2=0$ minimizes the return time at fixed $\gamma$. Let us clarify this point further.

\subsection{The critical line as an attractor of return time minimization}

To better understand the significance of the critical line, $\varepsilon^2=0$ , we consider the contour lines of $\tau$ and the gradient field
\begin{equation}
-\vec \nabla \tau = \left(- \frac{\partial \tau}{ \partial \gamma}, - \frac{\partial \tau}{ \partial \omega_0}\right)    
\end{equation}
in the ($\gamma, \omega_0$) plane (Fig.~\ref{fig:gradient}a).
The figure shows that the $\varepsilon^2=0$ line is an attractor for a gradient descent dynamics of $\tau$.  Although there are points outside the critical line that have a return time lower than some points on it, the gradient flows toward the critical line, so that a dynamic that tries to minimize the return time with local moves will end up along the line $\varepsilon^2=0$.
For example, the point $P_3$ in \figurename \ref{fig:gradient} (a) has a return time lower than $P_1$.
However the gradient flow does not take $P_1$ towards $P_3$; it rather takes both toward the point $P_2$.
To see this, consider the plane ($\gamma, \omega_0$) and call (A) the oscillating and (B) the non-oscillating regions (see \figurename \ref{fig:gradient} (a)).
For the critical line to be an attractor, the gradient lines in its neighborhood must point towards it.  Therefore one must have
\begin{align}
    - \frac{\partial \tau_{\text{\tiny A}}}{\partial \gamma} \bigg|_{\varepsilon^2 =0} &> - \frac{\partial \tau_{\text{\tiny A}}}{\partial \omega_0} \bigg|_{\varepsilon^2 =0}, \\
    - \frac{\partial \tau_{\text{\tiny B}}}{\partial \omega_0} \bigg|_{\varepsilon^2 =0} &> - \frac{\partial \tau_{\text{\tiny B}}}{\partial \gamma} \bigg|_{\varepsilon^2 =0}  .
\end{align}
In region A $\tau$ is given by \eqref{tauzero}, therefore
\begin{equation}
- \frac{\partial \tau_{\text{\tiny A}}}{\partial \gamma} \bigg|_{\varepsilon^2 =0} = \frac{1}{\gamma^2}\left[\log (\alpha \gamma) - 1\right] > 0 = - \frac{\partial \tau_{\text{\tiny A}}}{\partial \omega_0} \bigg|_{\varepsilon^2 =0}
\end{equation}
where $\alpha = u_0q/c$.  This condition is verified for $\gamma > e/\alpha$, i.e.\ outside the transparent zone.  In region B the return time is given by \eqref{taue}, so that the gradient is
\begin{align*}
- \frac{\partial \tau_{\text{\tiny B}}}{\partial \omega_0} \bigg|_{\varepsilon^2 =0} &= \frac{1}{2 \gamma^2} \left[\log(\alpha \gamma) + \log^2\left(\frac{u_0q\gamma}{c}\right) - \gamma^2 t_0^2\right], \\
- \frac{\partial \tau_{\text{\tiny B}}}{\partial \gamma} \bigg|_{\varepsilon^2 =0}  &=  \frac{1}{2 \gamma^2} \left[\log(\alpha \gamma) - \log^2(\alpha \gamma) + \gamma^2 t_0^2 - 2 \right]
\end{align*}
yielding
\begin{equation}
\log^2(\alpha \gamma) + 1 > \gamma^2 t_0^2 \, ,
\end{equation}
which is certainly verified if $\tau_0>0$, that is for $\log(\alpha \gamma) > \gamma t_0$.
This means that every point close to the part of the critical line with a positive return time, will flow to the critical line.
We conclude that the critical line is an attractor for the gradient dynamics of $\tau$. We may metaphorically view the (rather complicated) function $\tau(\gamma,\omega_0)$ as the main rib of a leaf, which is an attractor for a water drop (\figurename \ref{fig:gradient}c), although the situation here is a bit more complicated because of the non-trivial critical line structure.  

The maximum of $\tau$ on the critical line is a very special saddle point, because the flow field is not analytic at it: 
there is a separatrix that divides the basins of attraction of the transparent and of the opaque zone (\figurename \ref{fig:gradient}b).
So, depending on whether one starts to the left or to the right of the separatrix, the gradient flow will drive one to the transparent zone or in the opaque zone of the critical line, respectively.
We expect that a real system, and in particular the one we want to describe, lies close to the transparent zone: here the signal passes with weak attenuation and arrives still strong in every part of the system.

\section{How to look for evidence of speed waves in experimental data?}

\begin{figure*}[t!]
\centering
\includegraphics[width=.9\textwidth]{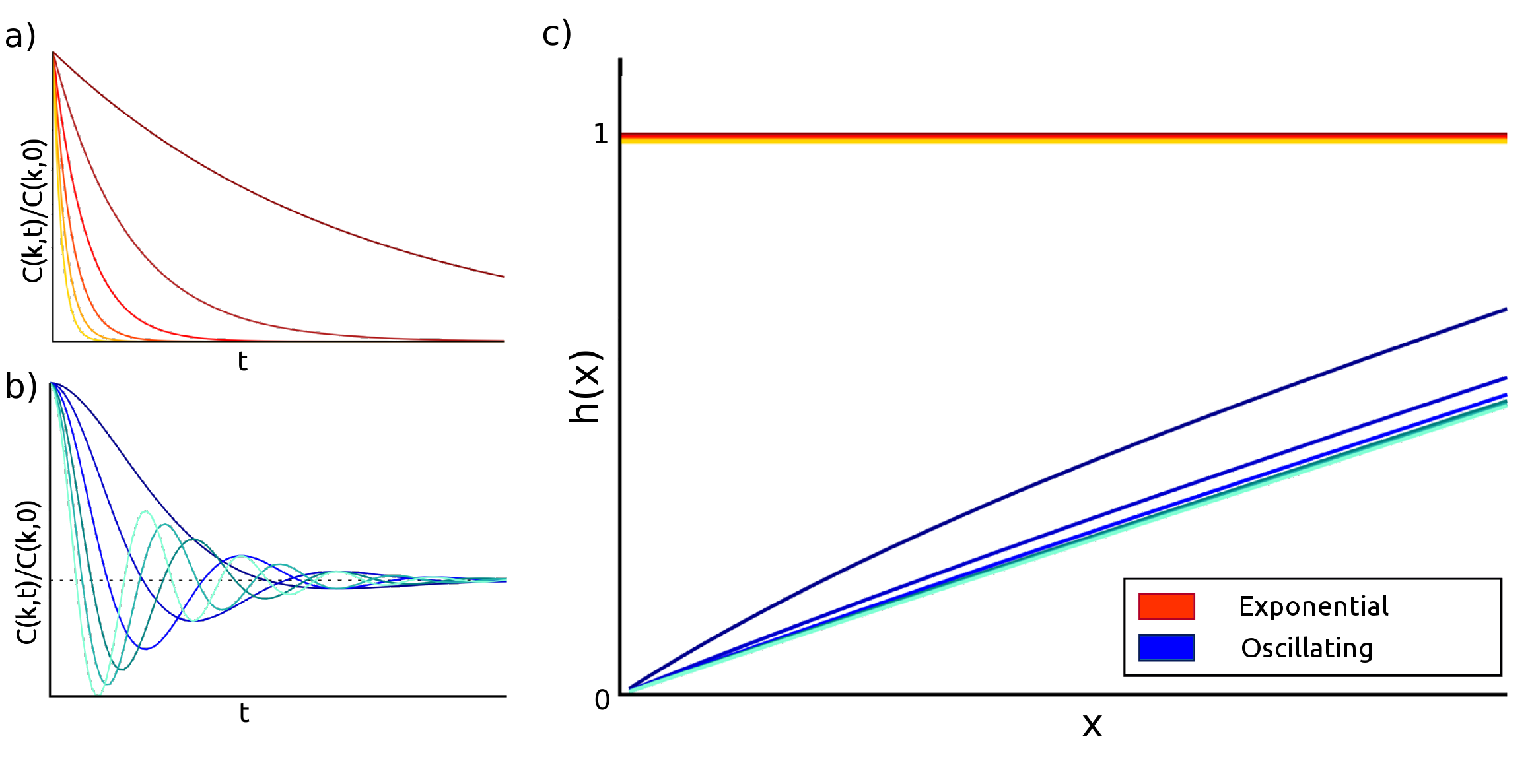}
\caption{(a) $C_k(t)$ in the non-oscillating regime $(\varepsilon^2 > 0)$ of the inertial dynamics for $k< k_0$ and for every value of $k$ in the Langevin dynamics.
The colors represent different values of $k$, ranging from dark red for small values of $k$, to yellow for high $k$.
(b) $C_k (t)$ in the inertial case for every value of $k$ if $\varepsilon^2 \ge 0$ and only for $k > k_0$ if $\varepsilon^2 >0$.
The colors represent different values of $k$, ranging from dark blue for small values of $k$, to aquamarine for high $k$.
(c) To better quantify the difference between inertial and non-inertial system we define the function $h(x)$ (see \cite{cavagna2017dynamic}) which has different form depending on whether the function is exponential-like or has a vanishing first derivative.}
\label{fig::corr}
\end{figure*}

To analyze how information propagates in a biological system directly, one has to observe an actual disturbance propagating in space and time.  However, naturally occurring disturbances may be relatively rare, and it is not always feasible to generate and artificial disturbance.  Another, indirect, way, is to analyze the spontaneous fluctuations of the system, that is to study dynamical correlations.
Indeed, qualitative features of the structure of the dynamical equations should leave identifiable traces in the shape of time correlations.  We have seen that the dynamic equations are quite different depending on the presence or absence of inertial terms; let us show how this is reflected in the dynamic correlation function.

\subsection{Spatio-temporal correlations}
We will focus on the intermediate scattering function, which is quite easy to compute at the experimental level \cite{cavagna2016spatio,cavagna2017dynamic},
\begin{equation}\begin{split}
C(k, t)&=\int d\textbf{x} \, e^{-i \textbf{k} \cdot \textbf{x}} \, C(r, t)\\
&= \int d \textbf{x} \, e^{-i \textbf{k} \cdot \textbf{x}} \langle u (\textbf{x},t_0)  u (\textbf{x} + r,t_0 + t) \rangle,
\end{split}\end{equation}
The spatio-temporal correlation function is a very useful tool, because its properties are entirely determined by the dispersion relation, which in turn mirrors the structure of the dynamical equation \cite{cavagna2016spatio}. Hence, one can infer from the behaviour of $C(k,t)$ a lot of information about the dynamics of a system. For the technical mathematical steps connecting the correlation function to the dispersion relation we refer the reader to \cite{cavagna2016spatio}.

In the case of Langevin dynamics (no speed waves), equations \eqref{langevin} and \eqref{zuppa}, $C(k,t)$ has the form (see \figurename \ref{fig::corr}a),
\begin{equation}
C(k,t)= \frac{2 T }{D k^2 + \omega_0} \, e^{-(D k^2 +\omega_0)\,t},
\end{equation}
where $T$ is the generalized temperature, and $D$ and $\omega_0$ have been defined in \eqref{mazza}. One can easily read the Langevin dispersion relation \eqref{zuppa} from the form of the correlation.

On the other hand, for the inertial dynamics of the speed wave equation ($\varepsilon^2<0$) the correlation function is given by (see \figurename \ref{fig::corr}b),
\begin{equation}\begin{split}
C(k,t)=& \frac{ \eta T }{\left(\frac{\tilde{J} k^2}{\mu} + \omega_0^2 \right)}\, e^{-\gamma t}\\
&\times\left\{ \frac{\sin \Big(\sqrt{\frac{\tilde{J} k^2}{\mu} - \varepsilon^2} \, t \Big)}{\sqrt{\frac{\tilde{J} k^2}{\mu} - \varepsilon^2}} + \frac{ \cos \Big(\sqrt{\frac{\tilde{J} k^2}{\mu} - \varepsilon^2} \, t \Big)}{\gamma} \right\},
\label{corrfull}
\end{split}\end{equation}
where $\tilde{J}= Ja^2 n_c$.  In the overdamped regime ($k< k_0 = \varepsilon/c$, $\varepsilon^2 >0$) the trigonometric functions must be replaced by the respective hyperbolic functions of argument ${\scriptstyle \left(\sqrt{\varepsilon^2 - (\tilde{J} k^2)/\mu } \, t \right)}$.
The form of the correlation function is considerably simpler than that of the full solution of the equation (Eq. \eqref{tel_I}, Eq. \eqref{tel_J}).
The correlation function refers only to a specific $k$ mode, while in the total solution all the modes are added giving rise to the Bessel functions.

\subsection{The fingerprint of inertial dynamics}

We see that while Langevin dynamics displays plain exponential relaxation, while inertial systems have a non-exponential oscillating correlation function. At first sight this may seem an obvious difference, very easy to detect from empirical data. However, the situation is more complex. First of all, empirical data typically derive from real $3D$ trajectories, which normally are not available for long times (the flock gets out of the field of view of our apparatus); if we have the correlation only for medium-short times it may be impossible to detect the oscillations, even if inertia (and therefore speed waves) are present. Secondly, if the system is close to critical damping, then there are {\it no} oscillations, even if speed waves are present! Hence, using oscillations as an empirical landmark of inertia and propagating waves is not a good idea. 

On the other hand, there is a feature of the correlation that is visible also for short times and that depends exclusively on the order of the dispersion relation (first vs second order), namely on the number of poles in the complex $\omega$ plane of the the correlation function. This feature is the first time derivative of the correlation for $t\to 0$ \cite{cavagna2017dynamic}. If the dispersion relation is of the first order, as in the Langevin case, then the derivative of the correlation in zero is finite, while if the dispersion relation is of the second order, as in the speed waves case, the derivative must go to zero. In order to quantitatively perform this analysis we can define the function,
\begin{equation}
h(x) = - \frac{1}{x} \log \left(\frac{C(x)}{C(0)} \right) \, , \,\, x \equiv t/t_k
\end{equation}
where $t_k$ is the characteristic time scale of the correlation, and study it in the interval $x\in[0,1]$, that is for times $t<t_k$.
For purely exponential relaxation $h(x) \rightarrow 1$ for $x \rightarrow 0$, while a flat time correlation gives $h(x) \rightarrow 0$ in the same limit (see \figurename \ref{fig::corr} (c)).
Once computed this function for real experimental data, if one has $h(x) \rightarrow 0$ then it is quite fair to say that the data have been generated by a dynamical equation that has inertial terms, therefore they are in a good agreement with the speed wave model. An experimental effort towards collecting this kind of data is currently under way.

\section{What kind of critical damping?}

The concept of critical damping in the context of collective behaviour was first introduced and studied in 2010 by Paley and coworkers \cite{paley2010critical}, which we now compare with our approach.

The first and most crucial difference between the two studies is that Paley and coworkers propose a $1d$ mathematical model directly 
for the {\it position}, rather than for the {\it speed}; hence, in \cite{paley2010critical}, the mechanism of imitation, typical of collective behaviour, amounts to imitating the position of the neighbours, rather than their speed. This is clearly visible in the mathematical expression of the model proposed in \cite{paley2010critical}, which is a second order dynamics for the positions,
\begin{equation}
\ddot{q}_i= \sum_{j \in n_i} -J (q_i - q_j - (i-j)q_0) - 2 \xi \sqrt{J}(\dot{q}_i - \dot{q}_j)
\label{panda}
\end{equation}
where $q_i$ is the position of the individual $i$, $\dot{q}_i$ its velocity, $J$ is the spring constant, $|i-j|q_0$ the rest length, while $2 \xi \sqrt{J}$ is the damping coefficient and $\xi>0$ (the inertia in \eqref{panda} is the normal mechanical mass, which is set to $1$). Instead of an anchoring term breaking the translational symmetry (translation in the speed in our case), \eqref{panda} has a linear damper connecting the particles. In order to make the comparison with our equation, we  rewrite \eqref{panda} in the continuous limit, 
\begin{equation}
\frac{\partial^2 q (\textbf{x},t)}{\partial t^2} + 2 \xi c \nabla^2 \frac{\partial q(\textbf{x},t)}{\partial t} -  c^2 \nabla^2 q(\textbf{x},t) = \zeta (\textbf{x},t) ,
\label{paley}
\end{equation}
where $q(\textbf{x},t)$ is the displacement field of the particles, $c= J a^2  n_c$, and $\zeta$ is a noise. This equation must be compared to our equation \eqref{full}. The dispersion relation associated to \eqref{paley} has the solution,
\begin{equation}
\omega = i \xi c k^2 \pm ck\sqrt{1 - \xi^2 k^2} \, .
\label{manga}
\end{equation}
For $k<1/\xi$ the frequency has a real part (propagating modes), while for $k >1/\xi$ the equation is overdamped; these two regimes are
separated by a {\it critical damping} value, $k=1/\xi$, and in \cite{paley2010critical} it is discussed how this edge is influenced by the connectivity of the network. Our dispersion relation \eqref{togotogo} can be rewritten as,
\begin{equation}
\omega = i \gamma \pm ck\sqrt{1 - \frac{\varepsilon^2}{c^2 k^2} } \ ,
\end{equation}
Here too there is a critical damping edge, $k = \varepsilon/c$, but its role is the {\it opposite} than in \eqref{manga}: the frequency is real for large $k$ and purely imaginary for low $k$.

To conclude, in the context of \cite{paley2010critical} critical damping does not concern the balance between inertia and dissipation, as in our study, but a transition between propagating and non-propagating modes in $k$ space. 
This type of definition of critical damping is not what impacts on the general solution of the dynamical equation in real space, which is found by summing over {\it all} $k$ modes: solutions \eqref{tel_I} and \eqref{tel_J} only depend on $\varepsilon$, that is on the balance between inertia and dissipation, and the critical damping value corresponds to $\varepsilon=0$, of which there is no analogue in the model developed in \cite{paley2010critical}

\section{Conclusions}

We proposed a new model for characterizing the propagation of speed fluctuations within highly polarized biological systems (flocks).
The resulting second order dynamical equation involves inertia, dissipation, interaction strength and a symmetry breaking term anchoring each individual to its physiological speed value. In general this equation has both underdamped and overdamped modes, 
giving rise to a complex structure of the general solution. However,  we found that along a certain line in the space of parameters, in particular when dissipation and inertia balance, the return time to the unperturbed state after a signal has passed is minimized. This is the critical damping line. We solved the equation exactly in one dimension and proved that the critical damping line is an attractor for a steepest descent dynamics of the return time. Finally, we proposed a method with which to assess, through an analysis of the experimental data, the validity of this model: by studying the dynamic correlations for speed it should be possible to verify the presence or absence of inertial terms in the dynamics and to refute/validate our model accordingly. 

Critical damping is quite a compelling concept at the biological level, especially in the case of speed waves. Let us consider a flock traveling unperturbed at a certain cruising speed. At some point an individual at the back of the flock detects a perturbation (as a predator), hence it changes its speed suddenly, giving rise to the propagation of a signal across the whole flock, which turns into a collective escaping maneuver. Clearly, after the signal has passed each individual will eventually go back to its physiological cruising speed. It seems reasonable to expect that this happens without oscillating back and forth around the cruising speed (this would seem utter nonsense), but also quite swiftly, in order to restore as quickly as possible the original dynamical state. If our theory is correct, such sensible way to go back to normal is achieved at critical damping. Experiments should easily detect whether inertial (second order) terms are present in the dynamics. Whether or not the dynamics is critically damped, though, will require to work out the different parameters, which  with our current experimental resolution seems harder, but not necessarily hopeless. Experimental efforts in this direction are under way.

\subsection*{Acknowledgements.}
This work was supported by IIT-Seed Artswarm and European Research Council Starting Grant 257126. AC thanks William Bialek for the interesting discussions on the subject of speed waves.

\bibliographystyle{apsrev4-1}
\bibliography{general_cobbs_bibliography_file}

\begin{thebibliography}{63}%
\makeatletter
\providecommand \@ifxundefined [1]{%
 \@ifx{#1\undefined}
}%
\providecommand \@ifnum [1]{%
 \ifnum #1\expandafter \@firstoftwo
 \else \expandafter \@secondoftwo
 \fi
}%
\providecommand \@ifx [1]{%
 \ifx #1\expandafter \@firstoftwo
 \else \expandafter \@secondoftwo
 \fi
}%
\providecommand \natexlab [1]{#1}%
\providecommand \enquote  [1]{``#1''}%
\providecommand \bibnamefont  [1]{#1}%
\providecommand \bibfnamefont [1]{#1}%
\providecommand \citenamefont [1]{#1}%
\providecommand \href@noop [0]{\@secondoftwo}%
\providecommand \href [0]{\begingroup \@sanitize@url \@href}%
\providecommand \@href[1]{\@@startlink{#1}\@@href}%
\providecommand \@@href[1]{\endgroup#1\@@endlink}%
\providecommand \@sanitize@url [0]{\catcode `\\12\catcode `\$12\catcode
  `\&12\catcode `\#12\catcode `\^12\catcode `\_12\catcode `\%12\relax}%
\providecommand \@@startlink[1]{}%
\providecommand \@@endlink[0]{}%
\providecommand \url  [0]{\begingroup\@sanitize@url \@url }%
\providecommand \@url [1]{\endgroup\@href {#1}{\urlprefix }}%
\providecommand \urlprefix  [0]{URL }%
\providecommand \Eprint [0]{\href }%
\providecommand \doibase [0]{http://dx.doi.org/}%
\providecommand \selectlanguage [0]{\@gobble}%
\providecommand \bibinfo  [0]{\@secondoftwo}%
\providecommand \bibfield  [0]{\@secondoftwo}%
\providecommand \translation [1]{[#1]}%
\providecommand \BibitemOpen [0]{}%
\providecommand \bibitemStop [0]{}%
\providecommand \bibitemNoStop [0]{.\EOS\space}%
\providecommand \EOS [0]{\spacefactor3000\relax}%
\providecommand \BibitemShut  [1]{\csname bibitem#1\endcsname}%
\let\auto@bib@innerbib\@empty
\bibitem [{\citenamefont {keng Ma}(2000)}]{Ma_book}%
  \BibitemOpen
  \bibfield  {author} {\bibinfo {author} {\bibfnamefont {S.}~\bibnamefont {keng
  Ma}},\ }\href@noop {} {\emph {\bibinfo {title} {Modern theory of critical
  phenomena}}},\ Advanced book classics\ (\bibinfo  {publisher} {Perseus Pub},\
  \bibinfo {year} {2000})\BibitemShut {NoStop}%
\bibitem [{\citenamefont {Parisi}(1988)}]{parisi_book}%
  \BibitemOpen
  \bibfield  {author} {\bibinfo {author} {\bibfnamefont {G.}~\bibnamefont
  {Parisi}},\ }\href {https://cds.cern.ch/record/111935} {\emph {\bibinfo
  {title} {{Statistical field theory}}}},\ Frontiers in Physics\ (\bibinfo
  {publisher} {Addison-Wesley},\ \bibinfo {address} {Redwood City, CA},\
  \bibinfo {year} {1988})\BibitemShut {NoStop}%
\bibitem [{\citenamefont {Sethna}(2006)}]{sethna2006statistical}%
  \BibitemOpen
  \bibfield  {author} {\bibinfo {author} {\bibfnamefont {J.}~\bibnamefont
  {Sethna}},\ }\href@noop {} {\emph {\bibinfo {title} {Statistical mechanics:
  entropy, order parameters, and complexity}}},\ Vol.~\bibinfo {volume} {14}\
  (\bibinfo  {publisher} {Oxford University Press},\ \bibinfo {year}
  {2006})\BibitemShut {NoStop}%
\bibitem [{\citenamefont {Sumpter}(2010)}]{sumpter_book}%
  \BibitemOpen
  \bibfield  {author} {\bibinfo {author} {\bibfnamefont {D.~J.}\ \bibnamefont
  {Sumpter}},\ }\href@noop {} {\emph {\bibinfo {title} {Collective animal
  behavior}}}\ (\bibinfo  {publisher} {Princeton University Press},\ \bibinfo
  {year} {2010})\BibitemShut {NoStop}%
\bibitem [{\citenamefont {Vicsek}\ and\ \citenamefont
  {Zafeiris}(2012)}]{vicsek_review}%
  \BibitemOpen
  \bibfield  {author} {\bibinfo {author} {\bibfnamefont {T.}~\bibnamefont
  {Vicsek}}\ and\ \bibinfo {author} {\bibfnamefont {A.}~\bibnamefont
  {Zafeiris}},\ }\href@noop {} {\bibfield  {journal} {\bibinfo  {journal}
  {Physics Reports}\ }\textbf {\bibinfo {volume} {517}},\ \bibinfo {pages} {71}
  (\bibinfo {year} {2012})}\BibitemShut {NoStop}%
\bibitem [{\citenamefont {Mora}\ and\ \citenamefont
  {Bialek}(2011)}]{mora+al_11}%
  \BibitemOpen
  \bibfield  {author} {\bibinfo {author} {\bibfnamefont {T.}~\bibnamefont
  {Mora}}\ and\ \bibinfo {author} {\bibfnamefont {W.}~\bibnamefont {Bialek}},\
  }\href {\doibase 10.1007/s10955-011-0229-4} {\bibfield  {journal} {\bibinfo
  {journal} {J Stat Phys}\ }\textbf {\bibinfo {volume} {144}},\ \bibinfo
  {pages} {268} (\bibinfo {year} {2011})}\BibitemShut {NoStop}%
\bibitem [{\citenamefont {Bialek}(2015)}]{bialek2015perspectives}%
  \BibitemOpen
  \bibfield  {author} {\bibinfo {author} {\bibfnamefont {W.}~\bibnamefont
  {Bialek}},\ }\href@noop {} {\bibfield  {journal} {\bibinfo  {journal} {arXiv
  preprint arXiv:1512.08954}\ } (\bibinfo {year} {2015})}\BibitemShut {NoStop}%
\bibitem [{\citenamefont {Stear}(1987)}]{stear1987control}%
  \BibitemOpen
  \bibfield  {author} {\bibinfo {author} {\bibfnamefont {E.~B.}\ \bibnamefont
  {Stear}},\ }in\ \href@noop {} {\emph {\bibinfo {booktitle} {Self-Organizing
  Systems}}}\ (\bibinfo  {publisher} {Springer},\ \bibinfo {year} {1987})\ pp.\
  \bibinfo {pages} {351--397}\BibitemShut {NoStop}%
\bibitem [{\citenamefont {Kube}\ and\ \citenamefont
  {Zhang}(1992)}]{kube1992collective}%
  \BibitemOpen
  \bibfield  {author} {\bibinfo {author} {\bibfnamefont {C.~R.}\ \bibnamefont
  {Kube}}\ and\ \bibinfo {author} {\bibfnamefont {H.}~\bibnamefont {Zhang}},\
  }in\ \href@noop {} {\emph {\bibinfo {booktitle} {Second International
  Conference on Simulation of Adaptive Behavior}}}\ (\bibinfo {year} {1992})\
  pp.\ \bibinfo {pages} {460--468}\BibitemShut {NoStop}%
\bibitem [{\citenamefont {Jadbabaie}\ \emph {et~al.}(2003)\citenamefont
  {Jadbabaie}, \citenamefont {Lin},\ and\ \citenamefont
  {Morse}}]{jadbabaie2003coordination}%
  \BibitemOpen
  \bibfield  {author} {\bibinfo {author} {\bibfnamefont {A.}~\bibnamefont
  {Jadbabaie}}, \bibinfo {author} {\bibfnamefont {J.}~\bibnamefont {Lin}}, \
  and\ \bibinfo {author} {\bibfnamefont {A.~S.}\ \bibnamefont {Morse}},\
  }\href@noop {} {\bibfield  {journal} {\bibinfo  {journal} {IEEE Transactions
  on automatic control}\ }\textbf {\bibinfo {volume} {48}},\ \bibinfo {pages}
  {988} (\bibinfo {year} {2003})}\BibitemShut {NoStop}%
\bibitem [{\citenamefont {Leonard}\ \emph {et~al.}(2007)\citenamefont
  {Leonard}, \citenamefont {Paley}, \citenamefont {Lekien}, \citenamefont
  {Sepulchre}, \citenamefont {Fratantoni},\ and\ \citenamefont
  {Davis}}]{leonard2007collective}%
  \BibitemOpen
  \bibfield  {author} {\bibinfo {author} {\bibfnamefont {N.~E.}\ \bibnamefont
  {Leonard}}, \bibinfo {author} {\bibfnamefont {D.~A.}\ \bibnamefont {Paley}},
  \bibinfo {author} {\bibfnamefont {F.}~\bibnamefont {Lekien}}, \bibinfo
  {author} {\bibfnamefont {R.}~\bibnamefont {Sepulchre}}, \bibinfo {author}
  {\bibfnamefont {D.~M.}\ \bibnamefont {Fratantoni}}, \ and\ \bibinfo {author}
  {\bibfnamefont {R.~E.}\ \bibnamefont {Davis}},\ }\href@noop {} {\bibfield
  {journal} {\bibinfo  {journal} {Proceedings of the IEEE}\ }\textbf {\bibinfo
  {volume} {95}},\ \bibinfo {pages} {48} (\bibinfo {year} {2007})}\BibitemShut
  {NoStop}%
\bibitem [{\citenamefont {Cavagna}\ \emph {et~al.}(2010)\citenamefont
  {Cavagna}, \citenamefont {Cimarelli}, \citenamefont {Giardina}, \citenamefont
  {Parisi}, \citenamefont {Santagati}, \citenamefont {Stefanini},\ and\
  \citenamefont {Viale}}]{cavagna+al_10}%
  \BibitemOpen
  \bibfield  {author} {\bibinfo {author} {\bibfnamefont {A.}~\bibnamefont
  {Cavagna}}, \bibinfo {author} {\bibfnamefont {A.}~\bibnamefont {Cimarelli}},
  \bibinfo {author} {\bibfnamefont {I.}~\bibnamefont {Giardina}}, \bibinfo
  {author} {\bibfnamefont {G.}~\bibnamefont {Parisi}}, \bibinfo {author}
  {\bibfnamefont {R.}~\bibnamefont {Santagati}}, \bibinfo {author}
  {\bibfnamefont {F.}~\bibnamefont {Stefanini}}, \ and\ \bibinfo {author}
  {\bibfnamefont {M.}~\bibnamefont {Viale}},\ }\href {\doibase
  10.1073/pnas.1005766107} {\bibfield  {journal} {\bibinfo  {journal} {Proc
  Natl Acad Sci USA}\ }\textbf {\bibinfo {volume} {107}},\ \bibinfo {pages}
  {11865} (\bibinfo {year} {2010})}\BibitemShut {NoStop}%
\bibitem [{\citenamefont {Attanasi}\ \emph
  {et~al.}(2014{\natexlab{a}})\citenamefont {Attanasi}, \citenamefont
  {Cavagna}, \citenamefont {Del~Castello}, \citenamefont {Giardina},
  \citenamefont {Melillo}, \citenamefont {Parisi}, \citenamefont {Pohl},
  \citenamefont {Rossaro}, \citenamefont {Shen}, \citenamefont {Silvestri}
  \emph {et~al.}}]{attanasi2014collective}%
  \BibitemOpen
  \bibfield  {author} {\bibinfo {author} {\bibfnamefont {A.}~\bibnamefont
  {Attanasi}}, \bibinfo {author} {\bibfnamefont {A.}~\bibnamefont {Cavagna}},
  \bibinfo {author} {\bibfnamefont {L.}~\bibnamefont {Del~Castello}}, \bibinfo
  {author} {\bibfnamefont {I.}~\bibnamefont {Giardina}}, \bibinfo {author}
  {\bibfnamefont {S.}~\bibnamefont {Melillo}}, \bibinfo {author} {\bibfnamefont
  {L.}~\bibnamefont {Parisi}}, \bibinfo {author} {\bibfnamefont
  {O.}~\bibnamefont {Pohl}}, \bibinfo {author} {\bibfnamefont {B.}~\bibnamefont
  {Rossaro}}, \bibinfo {author} {\bibfnamefont {E.}~\bibnamefont {Shen}},
  \bibinfo {author} {\bibfnamefont {E.}~\bibnamefont {Silvestri}},  \emph
  {et~al.},\ }\href@noop {} {\bibfield  {journal} {\bibinfo  {journal} {PLoS
  Comput Biol}\ }\textbf {\bibinfo {volume} {10}},\ \bibinfo {pages} {e1003697}
  (\bibinfo {year} {2014}{\natexlab{a}})}\BibitemShut {NoStop}%
\bibitem [{\citenamefont {Ginelli}\ \emph {et~al.}(2015)\citenamefont
  {Ginelli}, \citenamefont {Peruani}, \citenamefont {Pillot}, \citenamefont
  {Chat{\'e}}, \citenamefont {Theraulaz},\ and\ \citenamefont
  {Bon}}]{ginelli2015intermittent}%
  \BibitemOpen
  \bibfield  {author} {\bibinfo {author} {\bibfnamefont {F.}~\bibnamefont
  {Ginelli}}, \bibinfo {author} {\bibfnamefont {F.}~\bibnamefont {Peruani}},
  \bibinfo {author} {\bibfnamefont {M.-H.}\ \bibnamefont {Pillot}}, \bibinfo
  {author} {\bibfnamefont {H.}~\bibnamefont {Chat{\'e}}}, \bibinfo {author}
  {\bibfnamefont {G.}~\bibnamefont {Theraulaz}}, \ and\ \bibinfo {author}
  {\bibfnamefont {R.}~\bibnamefont {Bon}},\ }\href@noop {} {\bibfield
  {journal} {\bibinfo  {journal} {Proceedings of the National Academy of
  Sciences}\ }\textbf {\bibinfo {volume} {112}},\ \bibinfo {pages} {12729}
  (\bibinfo {year} {2015})}\BibitemShut {NoStop}%
\bibitem [{\citenamefont {Dombrowski}\ \emph {et~al.}(2004)\citenamefont
  {Dombrowski}, \citenamefont {Cisneros}, \citenamefont {Chatkaew},
  \citenamefont {Goldstein},\ and\ \citenamefont
  {Kessler}}]{dombrowski2004self}%
  \BibitemOpen
  \bibfield  {author} {\bibinfo {author} {\bibfnamefont {C.}~\bibnamefont
  {Dombrowski}}, \bibinfo {author} {\bibfnamefont {L.}~\bibnamefont
  {Cisneros}}, \bibinfo {author} {\bibfnamefont {S.}~\bibnamefont {Chatkaew}},
  \bibinfo {author} {\bibfnamefont {R.~E.}\ \bibnamefont {Goldstein}}, \ and\
  \bibinfo {author} {\bibfnamefont {J.~O.}\ \bibnamefont {Kessler}},\
  }\href@noop {} {\bibfield  {journal} {\bibinfo  {journal} {Physical Review
  Letters}\ }\textbf {\bibinfo {volume} {93}},\ \bibinfo {pages} {098103}
  (\bibinfo {year} {2004})}\BibitemShut {NoStop}%
\bibitem [{\citenamefont {Zhang}\ \emph {et~al.}(2010)\citenamefont {Zhang},
  \citenamefont {Be�er}, \citenamefont {Florin},\ and\ \citenamefont
  {Swinney}}]{zhang2010collective}%
  \BibitemOpen
  \bibfield  {author} {\bibinfo {author} {\bibfnamefont {H.-P.}\ \bibnamefont
  {Zhang}}, \bibinfo {author} {\bibfnamefont {A.}~\bibnamefont {Be�er}},
  \bibinfo {author} {\bibfnamefont {E.-L.}\ \bibnamefont {Florin}}, \ and\
  \bibinfo {author} {\bibfnamefont {H.~L.}\ \bibnamefont {Swinney}},\
  }\href@noop {} {\bibfield  {journal} {\bibinfo  {journal} {Proceedings of the
  National Academy of Sciences}\ }\textbf {\bibinfo {volume} {107}},\ \bibinfo
  {pages} {13626} (\bibinfo {year} {2010})}\BibitemShut {NoStop}%
\bibitem [{\citenamefont {Szabo}\ \emph {et~al.}(2006)\citenamefont {Szabo},
  \citenamefont {Sz{\"o}ll{\"o}si}, \citenamefont {G{\"o}nci}, \citenamefont
  {Jur{\'a}nyi}, \citenamefont {Selmeczi},\ and\ \citenamefont
  {Vicsek}}]{szabo2006phase}%
  \BibitemOpen
  \bibfield  {author} {\bibinfo {author} {\bibfnamefont {B.}~\bibnamefont
  {Szabo}}, \bibinfo {author} {\bibfnamefont {G.}~\bibnamefont
  {Sz{\"o}ll{\"o}si}}, \bibinfo {author} {\bibfnamefont {B.}~\bibnamefont
  {G{\"o}nci}}, \bibinfo {author} {\bibfnamefont {Z.}~\bibnamefont
  {Jur{\'a}nyi}}, \bibinfo {author} {\bibfnamefont {D.}~\bibnamefont
  {Selmeczi}}, \ and\ \bibinfo {author} {\bibfnamefont {T.}~\bibnamefont
  {Vicsek}},\ }\href@noop {} {\bibfield  {journal} {\bibinfo  {journal}
  {Physical Review E}\ }\textbf {\bibinfo {volume} {74}},\ \bibinfo {pages}
  {061908} (\bibinfo {year} {2006})}\BibitemShut {NoStop}%
\bibitem [{\citenamefont {Strandburg-Peshkin}\ \emph
  {et~al.}(2013)\citenamefont {Strandburg-Peshkin}, \citenamefont {Twomey},
  \citenamefont {Bode}, \citenamefont {Kao}, \citenamefont {Katz},
  \citenamefont {Ioannou}, \citenamefont {Rosenthal}, \citenamefont {Torney},
  \citenamefont {Wu}, \citenamefont {Levin} \emph {et~al.}}]{strandburg2013}%
  \BibitemOpen
  \bibfield  {author} {\bibinfo {author} {\bibfnamefont {A.}~\bibnamefont
  {Strandburg-Peshkin}}, \bibinfo {author} {\bibfnamefont {C.~R.}\ \bibnamefont
  {Twomey}}, \bibinfo {author} {\bibfnamefont {N.~W.}\ \bibnamefont {Bode}},
  \bibinfo {author} {\bibfnamefont {A.~B.}\ \bibnamefont {Kao}}, \bibinfo
  {author} {\bibfnamefont {Y.}~\bibnamefont {Katz}}, \bibinfo {author}
  {\bibfnamefont {C.~C.}\ \bibnamefont {Ioannou}}, \bibinfo {author}
  {\bibfnamefont {S.~B.}\ \bibnamefont {Rosenthal}}, \bibinfo {author}
  {\bibfnamefont {C.~J.}\ \bibnamefont {Torney}}, \bibinfo {author}
  {\bibfnamefont {H.~S.}\ \bibnamefont {Wu}}, \bibinfo {author} {\bibfnamefont
  {S.~A.}\ \bibnamefont {Levin}},  \emph {et~al.},\ }\href@noop {} {\bibfield
  {journal} {\bibinfo  {journal} {Current Biology}\ }\textbf {\bibinfo {volume}
  {23}},\ \bibinfo {pages} {R709} (\bibinfo {year} {2013})}\BibitemShut
  {NoStop}%
\bibitem [{\citenamefont {Goss}\ \emph {et~al.}(1989)\citenamefont {Goss},
  \citenamefont {Aron}, \citenamefont {Deneubourg},\ and\ \citenamefont
  {Pasteels}}]{goss1989self}%
  \BibitemOpen
  \bibfield  {author} {\bibinfo {author} {\bibfnamefont {S.}~\bibnamefont
  {Goss}}, \bibinfo {author} {\bibfnamefont {S.}~\bibnamefont {Aron}}, \bibinfo
  {author} {\bibfnamefont {J.-L.}\ \bibnamefont {Deneubourg}}, \ and\ \bibinfo
  {author} {\bibfnamefont {J.~M.}\ \bibnamefont {Pasteels}},\ }\href@noop {}
  {\bibfield  {journal} {\bibinfo  {journal} {Naturwissenschaften}\ }\textbf
  {\bibinfo {volume} {76}},\ \bibinfo {pages} {579} (\bibinfo {year}
  {1989})}\BibitemShut {NoStop}%
\bibitem [{\citenamefont {Cont}\ and\ \citenamefont
  {Bouchaud}(2000)}]{cont2000herd}%
  \BibitemOpen
  \bibfield  {author} {\bibinfo {author} {\bibfnamefont {R.}~\bibnamefont
  {Cont}}\ and\ \bibinfo {author} {\bibfnamefont {J.-P.}\ \bibnamefont
  {Bouchaud}},\ }\href@noop {} {\bibfield  {journal} {\bibinfo  {journal}
  {Macroeconomic dynamics}\ }\textbf {\bibinfo {volume} {4}},\ \bibinfo {pages}
  {170} (\bibinfo {year} {2000})}\BibitemShut {NoStop}%
\bibitem [{\citenamefont {Helbing}\ \emph {et~al.}(2001)\citenamefont
  {Helbing}, \citenamefont {Moln{\'a}r}, \citenamefont {Farkas},\ and\
  \citenamefont {Bolay}}]{helbing2001self}%
  \BibitemOpen
  \bibfield  {author} {\bibinfo {author} {\bibfnamefont {D.}~\bibnamefont
  {Helbing}}, \bibinfo {author} {\bibfnamefont {P.}~\bibnamefont {Moln{\'a}r}},
  \bibinfo {author} {\bibfnamefont {I.~J.}\ \bibnamefont {Farkas}}, \ and\
  \bibinfo {author} {\bibfnamefont {K.}~\bibnamefont {Bolay}},\ }\href@noop {}
  {\bibfield  {journal} {\bibinfo  {journal} {Environment and planning B:
  planning and design}\ }\textbf {\bibinfo {volume} {28}},\ \bibinfo {pages}
  {361} (\bibinfo {year} {2001})}\BibitemShut {NoStop}%
\bibitem [{\citenamefont {Giardina}(2008)}]{giardina2008cobag}%
  \BibitemOpen
  \bibfield  {author} {\bibinfo {author} {\bibfnamefont {I.}~\bibnamefont
  {Giardina}},\ }\href {\doibase 10.2976/1.2961038} {\bibfield  {journal}
  {\bibinfo  {journal} {HFSP Journal}\ }\textbf {\bibinfo {volume} {2}},\
  \bibinfo {pages} {205} (\bibinfo {year} {2008})},\ \bibinfo {note} {pMID:
  19404431}\BibitemShut {NoStop}%
\bibitem [{\citenamefont {Reynolds}(1987)}]{reynolds1987flocks}%
  \BibitemOpen
  \bibfield  {author} {\bibinfo {author} {\bibfnamefont {C.~W.}\ \bibnamefont
  {Reynolds}},\ }\href@noop {} {\bibfield  {journal} {\bibinfo  {journal} {ACM
  SIGGRAPH computer graphics}\ }\textbf {\bibinfo {volume} {21}},\ \bibinfo
  {pages} {25} (\bibinfo {year} {1987})}\BibitemShut {NoStop}%
\bibitem [{\citenamefont {Toner}\ and\ \citenamefont
  {Tu}(1998)}]{toner1998flocks}%
  \BibitemOpen
  \bibfield  {author} {\bibinfo {author} {\bibfnamefont {J.}~\bibnamefont
  {Toner}}\ and\ \bibinfo {author} {\bibfnamefont {Y.}~\bibnamefont {Tu}},\
  }\href@noop {} {\bibfield  {journal} {\bibinfo  {journal} {Physical review
  E}\ }\textbf {\bibinfo {volume} {58}},\ \bibinfo {pages} {4828} (\bibinfo
  {year} {1998})}\BibitemShut {NoStop}%
\bibitem [{\citenamefont {Bialek}\ \emph {et~al.}(2012)\citenamefont {Bialek},
  \citenamefont {Cavagna}, \citenamefont {Giardina}, \citenamefont {Mora},
  \citenamefont {Silvestri}, \citenamefont {Viale},\ and\ \citenamefont
  {Walczak}}]{bialek+al_12}%
  \BibitemOpen
  \bibfield  {author} {\bibinfo {author} {\bibfnamefont {W.}~\bibnamefont
  {Bialek}}, \bibinfo {author} {\bibfnamefont {A.}~\bibnamefont {Cavagna}},
  \bibinfo {author} {\bibfnamefont {I.}~\bibnamefont {Giardina}}, \bibinfo
  {author} {\bibfnamefont {T.}~\bibnamefont {Mora}}, \bibinfo {author}
  {\bibfnamefont {E.}~\bibnamefont {Silvestri}}, \bibinfo {author}
  {\bibfnamefont {M.}~\bibnamefont {Viale}}, \ and\ \bibinfo {author}
  {\bibfnamefont {A.~M.}\ \bibnamefont {Walczak}},\ }\href {\doibase
  10.1073/pnas.1118633109} {\bibfield  {journal} {\bibinfo  {journal} {Proc
  Natl Acad Sci USA}\ }\textbf {\bibinfo {volume} {109}},\ \bibinfo {pages}
  {4786} (\bibinfo {year} {2012})}\BibitemShut {NoStop}%
\bibitem [{\citenamefont {Hemelrijk}\ and\ \citenamefont
  {Hildenbrandt}(2012)}]{hemelrijk2012schools}%
  \BibitemOpen
  \bibfield  {author} {\bibinfo {author} {\bibfnamefont {C.~K.}\ \bibnamefont
  {Hemelrijk}}\ and\ \bibinfo {author} {\bibfnamefont {H.}~\bibnamefont
  {Hildenbrandt}},\ }\href@noop {} {\bibfield  {journal} {\bibinfo  {journal}
  {Interface focus}\ ,\ \bibinfo {pages} {rsfs20120025}} (\bibinfo {year}
  {2012})}\BibitemShut {NoStop}%
\bibitem [{\citenamefont {Cavagna}\ and\ \citenamefont
  {Giardina}(2014)}]{cavagna_review}%
  \BibitemOpen
  \bibfield  {author} {\bibinfo {author} {\bibfnamefont {A.}~\bibnamefont
  {Cavagna}}\ and\ \bibinfo {author} {\bibfnamefont {I.}~\bibnamefont
  {Giardina}},\ }\href@noop {} {\bibfield  {journal} {\bibinfo  {journal}
  {Annu. Rev. Condens. Matter Phys.}\ }\textbf {\bibinfo {volume} {5}},\
  \bibinfo {pages} {183} (\bibinfo {year} {2014})}\BibitemShut {NoStop}%
\bibitem [{\citenamefont {Vicsek}\ \emph {et~al.}(1995)\citenamefont {Vicsek},
  \citenamefont {Czir{\'o}k}, \citenamefont {Ben-Jacob}, \citenamefont
  {Cohen},\ and\ \citenamefont {Shochet}}]{vicsek+al_95}%
  \BibitemOpen
  \bibfield  {author} {\bibinfo {author} {\bibfnamefont {T.}~\bibnamefont
  {Vicsek}}, \bibinfo {author} {\bibfnamefont {A.}~\bibnamefont {Czir{\'o}k}},
  \bibinfo {author} {\bibfnamefont {E.}~\bibnamefont {Ben-Jacob}}, \bibinfo
  {author} {\bibfnamefont {I.}~\bibnamefont {Cohen}}, \ and\ \bibinfo {author}
  {\bibfnamefont {O.}~\bibnamefont {Shochet}},\ }\href@noop {} {\bibfield
  {journal} {\bibinfo  {journal} {Phys Rev Lett}\ }\textbf {\bibinfo {volume}
  {75}},\ \bibinfo {pages} {1226} (\bibinfo {year} {1995})}\BibitemShut
  {NoStop}%
\bibitem [{\citenamefont {Ramaswamy}(2010)}]{ramaswamy_review}%
  \BibitemOpen
  \bibfield  {author} {\bibinfo {author} {\bibfnamefont {S.}~\bibnamefont
  {Ramaswamy}},\ }\href@noop {} {\bibfield  {journal} {\bibinfo  {journal}
  {Annu. Rev. Condens. Matter Phys.}\ }\textbf {\bibinfo {volume} {1}},\
  \bibinfo {pages} {323} (\bibinfo {year} {2010})}\BibitemShut {NoStop}%
\bibitem [{\citenamefont {Marchetti}\ \emph {et~al.}(2013)\citenamefont
  {Marchetti}, \citenamefont {Joanny}, \citenamefont {Ramaswamy}, \citenamefont
  {Liverpool}, \citenamefont {Prost}, \citenamefont {Rao},\ and\ \citenamefont
  {Simha}}]{marchetti_review}%
  \BibitemOpen
  \bibfield  {author} {\bibinfo {author} {\bibfnamefont {M.}~\bibnamefont
  {Marchetti}}, \bibinfo {author} {\bibfnamefont {J.}~\bibnamefont {Joanny}},
  \bibinfo {author} {\bibfnamefont {S.}~\bibnamefont {Ramaswamy}}, \bibinfo
  {author} {\bibfnamefont {T.}~\bibnamefont {Liverpool}}, \bibinfo {author}
  {\bibfnamefont {J.}~\bibnamefont {Prost}}, \bibinfo {author} {\bibfnamefont
  {M.}~\bibnamefont {Rao}}, \ and\ \bibinfo {author} {\bibfnamefont {R.~A.}\
  \bibnamefont {Simha}},\ }\href@noop {} {\bibfield  {journal} {\bibinfo
  {journal} {Reviews of Modern Physics}\ }\textbf {\bibinfo {volume} {85}},\
  \bibinfo {pages} {1143} (\bibinfo {year} {2013})}\BibitemShut {NoStop}%
\bibitem [{\citenamefont {Ballerini}\ \emph
  {et~al.}(2008{\natexlab{a}})\citenamefont {Ballerini}, \citenamefont
  {Cabibbo}, \citenamefont {Candelier}, \citenamefont {Cavagna}, \citenamefont
  {Cisbani}, \citenamefont {Giardina}, \citenamefont {Orlandi}, \citenamefont
  {Parisi}, \citenamefont {Procaccini}, \citenamefont {Viale},\ and\
  \citenamefont {Zdravkovic}}]{ballerini+al_08b}%
  \BibitemOpen
  \bibfield  {author} {\bibinfo {author} {\bibfnamefont {M.}~\bibnamefont
  {Ballerini}}, \bibinfo {author} {\bibfnamefont {N.}~\bibnamefont {Cabibbo}},
  \bibinfo {author} {\bibfnamefont {R.}~\bibnamefont {Candelier}}, \bibinfo
  {author} {\bibfnamefont {A.}~\bibnamefont {Cavagna}}, \bibinfo {author}
  {\bibfnamefont {E.}~\bibnamefont {Cisbani}}, \bibinfo {author} {\bibfnamefont
  {I.}~\bibnamefont {Giardina}}, \bibinfo {author} {\bibfnamefont
  {A.}~\bibnamefont {Orlandi}}, \bibinfo {author} {\bibfnamefont
  {G.}~\bibnamefont {Parisi}}, \bibinfo {author} {\bibfnamefont
  {A.}~\bibnamefont {Procaccini}}, \bibinfo {author} {\bibfnamefont
  {M.}~\bibnamefont {Viale}}, \ and\ \bibinfo {author} {\bibfnamefont
  {V.}~\bibnamefont {Zdravkovic}},\ }\href {\doibase
  10.1016/j.anbehav.2008.02.004} {\bibfield  {journal} {\bibinfo  {journal}
  {Anim Behav}\ }\textbf {\bibinfo {volume} {76}},\ \bibinfo {pages} {201}
  (\bibinfo {year} {2008}{\natexlab{a}})}\BibitemShut {NoStop}%
\bibitem [{\citenamefont {Attanasi}\ \emph
  {et~al.}(2014{\natexlab{b}})\citenamefont {Attanasi}, \citenamefont
  {Cavagna}, \citenamefont {Del~Castello}, \citenamefont {Giardina},
  \citenamefont {Grigera}, \citenamefont {Jeli{\'c}}, \citenamefont {Melillo},
  \citenamefont {Parisi}, \citenamefont {Pohl}, \citenamefont {Shen} \emph
  {et~al.}}]{attanasi+al_14}%
  \BibitemOpen
  \bibfield  {author} {\bibinfo {author} {\bibfnamefont {A.}~\bibnamefont
  {Attanasi}}, \bibinfo {author} {\bibfnamefont {A.}~\bibnamefont {Cavagna}},
  \bibinfo {author} {\bibfnamefont {L.}~\bibnamefont {Del~Castello}}, \bibinfo
  {author} {\bibfnamefont {I.}~\bibnamefont {Giardina}}, \bibinfo {author}
  {\bibfnamefont {T.~S.}\ \bibnamefont {Grigera}}, \bibinfo {author}
  {\bibfnamefont {A.}~\bibnamefont {Jeli{\'c}}}, \bibinfo {author}
  {\bibfnamefont {S.}~\bibnamefont {Melillo}}, \bibinfo {author} {\bibfnamefont
  {L.}~\bibnamefont {Parisi}}, \bibinfo {author} {\bibfnamefont
  {O.}~\bibnamefont {Pohl}}, \bibinfo {author} {\bibfnamefont {E.}~\bibnamefont
  {Shen}},  \emph {et~al.},\ }\href@noop {} {\bibfield  {journal} {\bibinfo
  {journal} {Nature physics}\ }\textbf {\bibinfo {volume} {10}},\ \bibinfo
  {pages} {691} (\bibinfo {year} {2014}{\natexlab{b}})}\BibitemShut {NoStop}%
\bibitem [{\citenamefont {Cavagna}\ \emph
  {et~al.}(2015{\natexlab{a}})\citenamefont {Cavagna}, \citenamefont
  {Del~Castello}, \citenamefont {Giardina}, \citenamefont {Grigera},
  \citenamefont {Jelic}, \citenamefont {Melillo}, \citenamefont {Mora},
  \citenamefont {Parisi}, \citenamefont {Silvestri}, \citenamefont {Viale}
  \emph {et~al.}}]{cavagna+al_15}%
  \BibitemOpen
  \bibfield  {author} {\bibinfo {author} {\bibfnamefont {A.}~\bibnamefont
  {Cavagna}}, \bibinfo {author} {\bibfnamefont {L.}~\bibnamefont
  {Del~Castello}}, \bibinfo {author} {\bibfnamefont {I.}~\bibnamefont
  {Giardina}}, \bibinfo {author} {\bibfnamefont {T.}~\bibnamefont {Grigera}},
  \bibinfo {author} {\bibfnamefont {A.}~\bibnamefont {Jelic}}, \bibinfo
  {author} {\bibfnamefont {S.}~\bibnamefont {Melillo}}, \bibinfo {author}
  {\bibfnamefont {T.}~\bibnamefont {Mora}}, \bibinfo {author} {\bibfnamefont
  {L.}~\bibnamefont {Parisi}}, \bibinfo {author} {\bibfnamefont
  {E.}~\bibnamefont {Silvestri}}, \bibinfo {author} {\bibfnamefont
  {M.}~\bibnamefont {Viale}},  \emph {et~al.},\ }\href@noop {} {\bibfield
  {journal} {\bibinfo  {journal} {Journal of Statistical Physics}\ }\textbf
  {\bibinfo {volume} {158}},\ \bibinfo {pages} {601} (\bibinfo {year}
  {2015}{\natexlab{a}})}\BibitemShut {NoStop}%
\bibitem [{\citenamefont {Procaccini}\ \emph {et~al.}(2011)\citenamefont
  {Procaccini}, \citenamefont {Orlandi}, \citenamefont {Cavagna}, \citenamefont
  {Giardina}, \citenamefont {Zoratto}, \citenamefont {Santucci}, \citenamefont
  {Chiarotti}, \citenamefont {Hemelrijk}, \citenamefont {Alleva}, \citenamefont
  {Parisi},\ and\ \citenamefont {Carere}}]{procaccini+al2011}%
  \BibitemOpen
  \bibfield  {author} {\bibinfo {author} {\bibfnamefont {A.}~\bibnamefont
  {Procaccini}}, \bibinfo {author} {\bibfnamefont {A.}~\bibnamefont {Orlandi}},
  \bibinfo {author} {\bibfnamefont {A.}~\bibnamefont {Cavagna}}, \bibinfo
  {author} {\bibfnamefont {I.}~\bibnamefont {Giardina}}, \bibinfo {author}
  {\bibfnamefont {F.}~\bibnamefont {Zoratto}}, \bibinfo {author} {\bibfnamefont
  {D.}~\bibnamefont {Santucci}}, \bibinfo {author} {\bibfnamefont
  {F.}~\bibnamefont {Chiarotti}}, \bibinfo {author} {\bibfnamefont
  {C.}~\bibnamefont {Hemelrijk}}, \bibinfo {author} {\bibfnamefont
  {E.}~\bibnamefont {Alleva}}, \bibinfo {author} {\bibfnamefont
  {G.}~\bibnamefont {Parisi}}, \ and\ \bibinfo {author} {\bibfnamefont
  {C.}~\bibnamefont {Carere}},\ }\href {\doibase 10.1016/j.anbehav.2011.07.006}
  {\bibfield  {journal} {\bibinfo  {journal} {Animal Behavior}\ }\textbf
  {\bibinfo {volume} {82}},\ \bibinfo {pages} {759} (\bibinfo {year}
  {2011})}\BibitemShut {NoStop}%
\bibitem [{\citenamefont {Hemelrijk}\ \emph {et~al.}(2015)\citenamefont
  {Hemelrijk}, \citenamefont {van Zuidam},\ and\ \citenamefont
  {Hildenbrandt}}]{hemelrijk2015wave}%
  \BibitemOpen
  \bibfield  {author} {\bibinfo {author} {\bibfnamefont {C.~K.}\ \bibnamefont
  {Hemelrijk}}, \bibinfo {author} {\bibfnamefont {L.}~\bibnamefont {van
  Zuidam}}, \ and\ \bibinfo {author} {\bibfnamefont {H.}~\bibnamefont
  {Hildenbrandt}},\ }\href {\doibase 10.1007/s00265-015-1891-3} {\bibfield
  {journal} {\bibinfo  {journal} {Behavioral Ecology and Sociobiology}\
  }\textbf {\bibinfo {volume} {69}},\ \bibinfo {pages} {755} (\bibinfo {year}
  {2015})}\BibitemShut {NoStop}%
\bibitem [{\citenamefont {Tka{\v{c}}ik}\ and\ \citenamefont
  {Bialek}(2016)}]{tkavcik2016information}%
  \BibitemOpen
  \bibfield  {author} {\bibinfo {author} {\bibfnamefont {G.}~\bibnamefont
  {Tka{\v{c}}ik}}\ and\ \bibinfo {author} {\bibfnamefont {W.}~\bibnamefont
  {Bialek}},\ }\href@noop {} {\bibfield  {journal} {\bibinfo  {journal} {Annual
  Review of Condensed Matter Physics}\ }\textbf {\bibinfo {volume} {7}},\
  \bibinfo {pages} {89} (\bibinfo {year} {2016})}\BibitemShut {NoStop}%
\bibitem [{\citenamefont {Chat{\'e}}\ \emph {et~al.}(2008)\citenamefont
  {Chat{\'e}}, \citenamefont {Ginelli}, \citenamefont {Gr{\'e}goire},\ and\
  \citenamefont {Raynaud}}]{chate+al_08b}%
  \BibitemOpen
  \bibfield  {author} {\bibinfo {author} {\bibfnamefont {H.}~\bibnamefont
  {Chat{\'e}}}, \bibinfo {author} {\bibfnamefont {F.}~\bibnamefont {Ginelli}},
  \bibinfo {author} {\bibfnamefont {G.}~\bibnamefont {Gr{\'e}goire}}, \ and\
  \bibinfo {author} {\bibfnamefont {F.}~\bibnamefont {Raynaud}},\ }\href@noop
  {} {\bibfield  {journal} {\bibinfo  {journal} {Phys Rev E Stat Nonlin Soft
  Matter Phys}\ }\textbf {\bibinfo {volume} {77}},\ \bibinfo {pages} {046113}
  (\bibinfo {year} {2008})}\BibitemShut {NoStop}%
\bibitem [{\citenamefont {Toner}\ and\ \citenamefont {Tu}(1995)}]{toner+al_95}%
  \BibitemOpen
  \bibfield  {author} {\bibinfo {author} {\bibfnamefont {J.}~\bibnamefont
  {Toner}}\ and\ \bibinfo {author} {\bibfnamefont {Y.}~\bibnamefont {Tu}},\
  }\href@noop {} {\bibfield  {journal} {\bibinfo  {journal} {Phys Rev Lett}\
  }\textbf {\bibinfo {volume} {75}},\ \bibinfo {pages} {4326} (\bibinfo {year}
  {1995})}\BibitemShut {NoStop}%
\bibitem [{\citenamefont {Tu}\ \emph {et~al.}(1998)\citenamefont {Tu},
  \citenamefont {Toner},\ and\ \citenamefont {Ulm}}]{toner+al_98}%
  \BibitemOpen
  \bibfield  {author} {\bibinfo {author} {\bibfnamefont {Y.}~\bibnamefont
  {Tu}}, \bibinfo {author} {\bibfnamefont {J.}~\bibnamefont {Toner}}, \ and\
  \bibinfo {author} {\bibfnamefont {M.}~\bibnamefont {Ulm}},\ }\href {\doibase
  10.1103/PhysRevLett.80.4819} {\bibfield  {journal} {\bibinfo  {journal}
  {Phys. Rev. Lett.}\ }\textbf {\bibinfo {volume} {80}},\ \bibinfo {pages}
  {4819} (\bibinfo {year} {1998})}\BibitemShut {NoStop}%
\bibitem [{\citenamefont {Toner}\ \emph {et~al.}(2005)\citenamefont {Toner},
  \citenamefont {Tu},\ and\ \citenamefont {Ramaswamy}}]{toner_review}%
  \BibitemOpen
  \bibfield  {author} {\bibinfo {author} {\bibfnamefont {J.}~\bibnamefont
  {Toner}}, \bibinfo {author} {\bibfnamefont {Y.}~\bibnamefont {Tu}}, \ and\
  \bibinfo {author} {\bibfnamefont {S.}~\bibnamefont {Ramaswamy}},\ }\href@noop
  {} {\bibfield  {journal} {\bibinfo  {journal} {Annals of Physics}\ }\textbf
  {\bibinfo {volume} {318}},\ \bibinfo {pages} {170} (\bibinfo {year}
  {2005})}\BibitemShut {NoStop}%
\bibitem [{\citenamefont {Bialek}\ \emph {et~al.}(2014)\citenamefont {Bialek},
  \citenamefont {Cavagna}, \citenamefont {Giardina}, \citenamefont {Mora},
  \citenamefont {Pohl}, \citenamefont {Silvestri}, \citenamefont {Viale},\ and\
  \citenamefont {Walczak}}]{bialek+al_14}%
  \BibitemOpen
  \bibfield  {author} {\bibinfo {author} {\bibfnamefont {W.}~\bibnamefont
  {Bialek}}, \bibinfo {author} {\bibfnamefont {A.}~\bibnamefont {Cavagna}},
  \bibinfo {author} {\bibfnamefont {I.}~\bibnamefont {Giardina}}, \bibinfo
  {author} {\bibfnamefont {T.}~\bibnamefont {Mora}}, \bibinfo {author}
  {\bibfnamefont {O.}~\bibnamefont {Pohl}}, \bibinfo {author} {\bibfnamefont
  {E.}~\bibnamefont {Silvestri}}, \bibinfo {author} {\bibfnamefont
  {M.}~\bibnamefont {Viale}}, \ and\ \bibinfo {author} {\bibfnamefont {A.~M.}\
  \bibnamefont {Walczak}},\ }\href@noop {} {\bibfield  {journal} {\bibinfo
  {journal} {Proceedings of the National Academy of Sciences}\ }\textbf
  {\bibinfo {volume} {111}},\ \bibinfo {pages} {7212} (\bibinfo {year}
  {2014})}\BibitemShut {NoStop}%
\bibitem [{\citenamefont {Hemelrijk}\ and\ \citenamefont
  {Hildenbrandt}(2015)}]{hemelrijk2015scale}%
  \BibitemOpen
  \bibfield  {author} {\bibinfo {author} {\bibfnamefont {C.~K.}\ \bibnamefont
  {Hemelrijk}}\ and\ \bibinfo {author} {\bibfnamefont {H.}~\bibnamefont
  {Hildenbrandt}},\ }\href@noop {} {\bibfield  {journal} {\bibinfo  {journal}
  {Journal of Statistical Physics}\ }\textbf {\bibinfo {volume} {158}},\
  \bibinfo {pages} {563} (\bibinfo {year} {2015})}\BibitemShut {NoStop}%
\bibitem [{\citenamefont {Peruani}\ and\ \citenamefont
  {Morelli}(2007)}]{peruani2007self}%
  \BibitemOpen
  \bibfield  {author} {\bibinfo {author} {\bibfnamefont {F.}~\bibnamefont
  {Peruani}}\ and\ \bibinfo {author} {\bibfnamefont {L.~G.}\ \bibnamefont
  {Morelli}},\ }\href@noop {} {\bibfield  {journal} {\bibinfo  {journal}
  {Physical review letters}\ }\textbf {\bibinfo {volume} {99}},\ \bibinfo
  {pages} {010602} (\bibinfo {year} {2007})}\BibitemShut {NoStop}%
\bibitem [{\citenamefont {Smirnov}(1964)}]{smirnov1964course}%
  \BibitemOpen
  \bibfield  {author} {\bibinfo {author} {\bibfnamefont {V.~I.}\ \bibnamefont
  {Smirnov}},\ }\href@noop {} {\emph {\bibinfo {title} {A Course of Higher
  Mathematics: Vol. 2, Advanced Calculus}}}\ (\bibinfo  {publisher} {Pergamon
  Press},\ \bibinfo {year} {1964})\BibitemShut {NoStop}%
\bibitem [{\citenamefont {Webster}\ and\ \citenamefont
  {Plimpton}(2016)}]{webster2016partial}%
  \BibitemOpen
  \bibfield  {author} {\bibinfo {author} {\bibfnamefont {A.~G.}\ \bibnamefont
  {Webster}}\ and\ \bibinfo {author} {\bibfnamefont {S.~J.}\ \bibnamefont
  {Plimpton}},\ }\href@noop {} {\emph {\bibinfo {title} {Partial differential
  equations of mathematical physics}}}\ (\bibinfo  {publisher} {Courier Dover
  Publications},\ \bibinfo {year} {2016})\BibitemShut {NoStop}%
\bibitem [{\citenamefont {Magnusson}\ \emph {et~al.}(2000)\citenamefont
  {Magnusson}, \citenamefont {Weisshaar}, \citenamefont {Tripathi},\ and\
  \citenamefont {Alexander}}]{magnusson2000transmission}%
  \BibitemOpen
  \bibfield  {author} {\bibinfo {author} {\bibfnamefont {P.~C.}\ \bibnamefont
  {Magnusson}}, \bibinfo {author} {\bibfnamefont {A.}~\bibnamefont
  {Weisshaar}}, \bibinfo {author} {\bibfnamefont {V.~K.}\ \bibnamefont
  {Tripathi}}, \ and\ \bibinfo {author} {\bibfnamefont {G.~C.}\ \bibnamefont
  {Alexander}},\ }\href@noop {} {\emph {\bibinfo {title} {Transmission lines
  and wave propagation}}}\ (\bibinfo  {publisher} {CRC Press},\ \bibinfo {year}
  {2000})\BibitemShut {NoStop}%
\bibitem [{\citenamefont {Ballerini}\ \emph
  {et~al.}(2008{\natexlab{b}})\citenamefont {Ballerini}, \citenamefont
  {Cabibbo}, \citenamefont {Candelier}, \citenamefont {Cavagna}, \citenamefont
  {Cisbani}, \citenamefont {Giardina}, \citenamefont {Lecomte}, \citenamefont
  {Orlandi}, \citenamefont {Parisi}, \citenamefont {Procaccini} \emph
  {et~al.}}]{ballerini+al_08}%
  \BibitemOpen
  \bibfield  {author} {\bibinfo {author} {\bibfnamefont {M.}~\bibnamefont
  {Ballerini}}, \bibinfo {author} {\bibfnamefont {N.}~\bibnamefont {Cabibbo}},
  \bibinfo {author} {\bibfnamefont {R.}~\bibnamefont {Candelier}}, \bibinfo
  {author} {\bibfnamefont {A.}~\bibnamefont {Cavagna}}, \bibinfo {author}
  {\bibfnamefont {E.}~\bibnamefont {Cisbani}}, \bibinfo {author} {\bibfnamefont
  {I.}~\bibnamefont {Giardina}}, \bibinfo {author} {\bibfnamefont
  {V.}~\bibnamefont {Lecomte}}, \bibinfo {author} {\bibfnamefont
  {A.}~\bibnamefont {Orlandi}}, \bibinfo {author} {\bibfnamefont
  {G.}~\bibnamefont {Parisi}}, \bibinfo {author} {\bibfnamefont
  {A.}~\bibnamefont {Procaccini}},  \emph {et~al.},\ }\href@noop {} {\bibfield
  {journal} {\bibinfo  {journal} {Proceedings of the national academy of
  sciences}\ }\textbf {\bibinfo {volume} {105}},\ \bibinfo {pages} {1232}
  (\bibinfo {year} {2008}{\natexlab{b}})}\BibitemShut {NoStop}%
\bibitem [{\citenamefont {Ginelli}\ and\ \citenamefont
  {Chat{\'e}}(2010)}]{ginelli+al_10}%
  \BibitemOpen
  \bibfield  {author} {\bibinfo {author} {\bibfnamefont {F.}~\bibnamefont
  {Ginelli}}\ and\ \bibinfo {author} {\bibfnamefont {H.}~\bibnamefont
  {Chat{\'e}}},\ }\href@noop {} {\bibfield  {journal} {\bibinfo  {journal}
  {Phys Rev Lett}\ }\textbf {\bibinfo {volume} {105}},\ \bibinfo {pages}
  {168103} (\bibinfo {year} {2010})}\BibitemShut {NoStop}%
\bibitem [{\citenamefont {Cavagna}\ \emph {et~al.}(2014)\citenamefont
  {Cavagna}, \citenamefont {Giardina}, \citenamefont {Ginelli}, \citenamefont
  {Mora}, \citenamefont {Piovani}, \citenamefont {Tavarone},\ and\
  \citenamefont {Walczak}}]{cavagna2014dynamical}%
  \BibitemOpen
  \bibfield  {author} {\bibinfo {author} {\bibfnamefont {A.}~\bibnamefont
  {Cavagna}}, \bibinfo {author} {\bibfnamefont {I.}~\bibnamefont {Giardina}},
  \bibinfo {author} {\bibfnamefont {F.}~\bibnamefont {Ginelli}}, \bibinfo
  {author} {\bibfnamefont {T.}~\bibnamefont {Mora}}, \bibinfo {author}
  {\bibfnamefont {D.}~\bibnamefont {Piovani}}, \bibinfo {author} {\bibfnamefont
  {R.}~\bibnamefont {Tavarone}}, \ and\ \bibinfo {author} {\bibfnamefont
  {A.~M.}\ \bibnamefont {Walczak}},\ }\href@noop {} {\bibfield  {journal}
  {\bibinfo  {journal} {Physical Review E}\ }\textbf {\bibinfo {volume} {89}},\
  \bibinfo {pages} {042707} (\bibinfo {year} {2014})}\BibitemShut {NoStop}%
\bibitem [{\citenamefont {Zwanzig}(2001)}]{zwanzig_book}%
  \BibitemOpen
  \bibfield  {author} {\bibinfo {author} {\bibfnamefont {R.}~\bibnamefont
  {Zwanzig}},\ }\href@noop {} {\emph {\bibinfo {title} {Nonequilibrium
  statistical mechanics}}}\ (\bibinfo  {publisher} {Oxford University Press,
  USA},\ \bibinfo {year} {2001})\BibitemShut {NoStop}%
\bibitem [{\citenamefont {Fodor}\ \emph {et~al.}(2016)\citenamefont {Fodor},
  \citenamefont {Nardini}, \citenamefont {Cates}, \citenamefont {Tailleur},
  \citenamefont {Visco},\ and\ \citenamefont {van Wijland}}]{fodor2016far}%
  \BibitemOpen
  \bibfield  {author} {\bibinfo {author} {\bibfnamefont {{\'E}.}~\bibnamefont
  {Fodor}}, \bibinfo {author} {\bibfnamefont {C.}~\bibnamefont {Nardini}},
  \bibinfo {author} {\bibfnamefont {M.~E.}\ \bibnamefont {Cates}}, \bibinfo
  {author} {\bibfnamefont {J.}~\bibnamefont {Tailleur}}, \bibinfo {author}
  {\bibfnamefont {P.}~\bibnamefont {Visco}}, \ and\ \bibinfo {author}
  {\bibfnamefont {F.}~\bibnamefont {van Wijland}},\ }\href@noop {} {\bibfield
  {journal} {\bibinfo  {journal} {Physical Review Letters}\ }\textbf {\bibinfo
  {volume} {117}},\ \bibinfo {pages} {038103} (\bibinfo {year}
  {2016})}\BibitemShut {NoStop}%
\bibitem [{\citenamefont {Loi}\ \emph {et~al.}(2008)\citenamefont {Loi},
  \citenamefont {Mossa},\ and\ \citenamefont {Cugliandolo}}]{loi2008effective}%
  \BibitemOpen
  \bibfield  {author} {\bibinfo {author} {\bibfnamefont {D.}~\bibnamefont
  {Loi}}, \bibinfo {author} {\bibfnamefont {S.}~\bibnamefont {Mossa}}, \ and\
  \bibinfo {author} {\bibfnamefont {L.~F.}\ \bibnamefont {Cugliandolo}},\
  }\href@noop {} {\bibfield  {journal} {\bibinfo  {journal} {Physical Review
  E}\ }\textbf {\bibinfo {volume} {77}},\ \bibinfo {pages} {051111} (\bibinfo
  {year} {2008})}\BibitemShut {NoStop}%
\bibitem [{\citenamefont {Marconi}\ and\ \citenamefont
  {Maggi}(2015)}]{marconi2015towards}%
  \BibitemOpen
  \bibfield  {author} {\bibinfo {author} {\bibfnamefont {U.~M.~B.}\
  \bibnamefont {Marconi}}\ and\ \bibinfo {author} {\bibfnamefont
  {C.}~\bibnamefont {Maggi}},\ }\href@noop {} {\bibfield  {journal} {\bibinfo
  {journal} {Soft matter}\ }\textbf {\bibinfo {volume} {11}},\ \bibinfo {pages}
  {8768} (\bibinfo {year} {2015})}\BibitemShut {NoStop}%
\bibitem [{\citenamefont {Cavagna}\ \emph {et~al.}(2013)\citenamefont
  {Cavagna}, \citenamefont {Queir{\'o}s}, \citenamefont {Giardina},
  \citenamefont {Stefanini},\ and\ \citenamefont {Viale}}]{cavagna+al_13}%
  \BibitemOpen
  \bibfield  {author} {\bibinfo {author} {\bibfnamefont {A.}~\bibnamefont
  {Cavagna}}, \bibinfo {author} {\bibfnamefont {S.~M.~D.}\ \bibnamefont
  {Queir{\'o}s}}, \bibinfo {author} {\bibfnamefont {I.}~\bibnamefont
  {Giardina}}, \bibinfo {author} {\bibfnamefont {F.}~\bibnamefont {Stefanini}},
  \ and\ \bibinfo {author} {\bibfnamefont {M.}~\bibnamefont {Viale}},\ }\href
  {\doibase 10.1098/rspb.2012.2484} {\bibfield  {journal} {\bibinfo  {journal}
  {Proc Biol Sci}\ }\textbf {\bibinfo {volume} {280}},\ \bibinfo {pages}
  {20122484} (\bibinfo {year} {2013})}\BibitemShut {NoStop}%
\bibitem [{\citenamefont {Mora}\ \emph {et~al.}(2016)\citenamefont {Mora},
  \citenamefont {Walczak}, \citenamefont {Del~Castello}, \citenamefont
  {Ginelli}, \citenamefont {Melillo}, \citenamefont {Parisi}, \citenamefont
  {Viale}, \citenamefont {Cavagna},\ and\ \citenamefont
  {Giardina}}]{mora2016local}%
  \BibitemOpen
  \bibfield  {author} {\bibinfo {author} {\bibfnamefont {T.}~\bibnamefont
  {Mora}}, \bibinfo {author} {\bibfnamefont {A.~M.}\ \bibnamefont {Walczak}},
  \bibinfo {author} {\bibfnamefont {L.}~\bibnamefont {Del~Castello}}, \bibinfo
  {author} {\bibfnamefont {F.}~\bibnamefont {Ginelli}}, \bibinfo {author}
  {\bibfnamefont {S.}~\bibnamefont {Melillo}}, \bibinfo {author} {\bibfnamefont
  {L.}~\bibnamefont {Parisi}}, \bibinfo {author} {\bibfnamefont
  {M.}~\bibnamefont {Viale}}, \bibinfo {author} {\bibfnamefont
  {A.}~\bibnamefont {Cavagna}}, \ and\ \bibinfo {author} {\bibfnamefont
  {I.}~\bibnamefont {Giardina}},\ }\href@noop {} {\bibfield  {journal}
  {\bibinfo  {journal} {Nature Physics}\ }\textbf {\bibinfo {volume} {12}},\
  \bibinfo {pages} {1153} (\bibinfo {year} {2016})}\BibitemShut {NoStop}%
\bibitem [{\citenamefont {Cavagna}\ \emph
  {et~al.}(2015{\natexlab{b}})\citenamefont {Cavagna}, \citenamefont
  {Giardina}, \citenamefont {Grigera}, \citenamefont {Jelic}, \citenamefont
  {Levine}, \citenamefont {Ramaswamy},\ and\ \citenamefont
  {Viale}}]{cavagna2015silent}%
  \BibitemOpen
  \bibfield  {author} {\bibinfo {author} {\bibfnamefont {A.}~\bibnamefont
  {Cavagna}}, \bibinfo {author} {\bibfnamefont {I.}~\bibnamefont {Giardina}},
  \bibinfo {author} {\bibfnamefont {T.~S.}\ \bibnamefont {Grigera}}, \bibinfo
  {author} {\bibfnamefont {A.}~\bibnamefont {Jelic}}, \bibinfo {author}
  {\bibfnamefont {D.}~\bibnamefont {Levine}}, \bibinfo {author} {\bibfnamefont
  {S.}~\bibnamefont {Ramaswamy}}, \ and\ \bibinfo {author} {\bibfnamefont
  {M.}~\bibnamefont {Viale}},\ }\href@noop {} {\bibfield  {journal} {\bibinfo
  {journal} {Physical review letters}\ }\textbf {\bibinfo {volume} {114}},\
  \bibinfo {pages} {218101} (\bibinfo {year} {2015}{\natexlab{b}})}\BibitemShut
  {NoStop}%
\bibitem [{\citenamefont {Huang}(2009)}]{huang2009introduction}%
  \BibitemOpen
  \bibfield  {author} {\bibinfo {author} {\bibfnamefont {K.}~\bibnamefont
  {Huang}},\ }\href@noop {} {\emph {\bibinfo {title} {Introduction to
  statistical physics}}}\ (\bibinfo  {publisher} {CRC press},\ \bibinfo {year}
  {2009})\BibitemShut {NoStop}%
\bibitem [{\citenamefont {Strauss}(1992)}]{strauss1992partial}%
  \BibitemOpen
  \bibfield  {author} {\bibinfo {author} {\bibfnamefont {W.~A.}\ \bibnamefont
  {Strauss}},\ }\href@noop {} {\emph {\bibinfo {title} {Partial differential
  equations}}},\ Vol.~\bibinfo {volume} {92}\ (\bibinfo  {publisher} {Wiley New
  York},\ \bibinfo {year} {1992})\BibitemShut {NoStop}%
\bibitem [{\citenamefont {Morse}\ and\ \citenamefont
  {Feshbach}(1946)}]{morse1946methods}%
  \BibitemOpen
  \bibfield  {author} {\bibinfo {author} {\bibfnamefont {P.~M.}\ \bibnamefont
  {Morse}}\ and\ \bibinfo {author} {\bibfnamefont {H.}~\bibnamefont
  {Feshbach}},\ }\href@noop {} {\emph {\bibinfo {title} {Methods of theoretical
  physics}}}\ (\bibinfo  {publisher} {Technology Press},\ \bibinfo {year}
  {1946})\BibitemShut {NoStop}%
\bibitem [{\citenamefont {Taylor}(2005)}]{taylor2005classical}%
  \BibitemOpen
  \bibfield  {author} {\bibinfo {author} {\bibfnamefont {J.~R.}\ \bibnamefont
  {Taylor}},\ }\href@noop {} {\emph {\bibinfo {title} {Classical mechanics}}}\
  (\bibinfo  {publisher} {University Science Books},\ \bibinfo {year}
  {2005})\BibitemShut {NoStop}%
\bibitem [{\citenamefont {Cavagna}\ \emph {et~al.}(2017)\citenamefont
  {Cavagna}, \citenamefont {Conti}, \citenamefont {Creato}, \citenamefont
  {Del~Castello}, \citenamefont {Giardina}, \citenamefont {Grigera},
  \citenamefont {Melillo}, \citenamefont {Parisi},\ and\ \citenamefont
  {Viale}}]{cavagna2017dynamic}%
  \BibitemOpen
  \bibfield  {author} {\bibinfo {author} {\bibfnamefont {A.}~\bibnamefont
  {Cavagna}}, \bibinfo {author} {\bibfnamefont {D.}~\bibnamefont {Conti}},
  \bibinfo {author} {\bibfnamefont {C.}~\bibnamefont {Creato}}, \bibinfo
  {author} {\bibfnamefont {L.}~\bibnamefont {Del~Castello}}, \bibinfo {author}
  {\bibfnamefont {I.}~\bibnamefont {Giardina}}, \bibinfo {author}
  {\bibfnamefont {T.~S.}\ \bibnamefont {Grigera}}, \bibinfo {author}
  {\bibfnamefont {S.}~\bibnamefont {Melillo}}, \bibinfo {author} {\bibfnamefont
  {L.}~\bibnamefont {Parisi}}, \ and\ \bibinfo {author} {\bibfnamefont
  {M.}~\bibnamefont {Viale}},\ }\href@noop {} {\bibfield  {journal} {\bibinfo
  {journal} {Nature Physics}\ } (\bibinfo {year} {2017})}\BibitemShut {NoStop}%
\bibitem [{\citenamefont {Cavagna}\ \emph {et~al.}(2016)\citenamefont
  {Cavagna}, \citenamefont {Conti}, \citenamefont {Giardina}, \citenamefont
  {Grigera}, \citenamefont {Melillo},\ and\ \citenamefont
  {Viale}}]{cavagna2016spatio}%
  \BibitemOpen
  \bibfield  {author} {\bibinfo {author} {\bibfnamefont {A.}~\bibnamefont
  {Cavagna}}, \bibinfo {author} {\bibfnamefont {D.}~\bibnamefont {Conti}},
  \bibinfo {author} {\bibfnamefont {I.}~\bibnamefont {Giardina}}, \bibinfo
  {author} {\bibfnamefont {T.~S.}\ \bibnamefont {Grigera}}, \bibinfo {author}
  {\bibfnamefont {S.}~\bibnamefont {Melillo}}, \ and\ \bibinfo {author}
  {\bibfnamefont {M.}~\bibnamefont {Viale}},\ }\href@noop {} {\bibfield
  {journal} {\bibinfo  {journal} {Physical Biology}\ }\textbf {\bibinfo
  {volume} {13}},\ \bibinfo {pages} {065001} (\bibinfo {year}
  {2016})}\BibitemShut {NoStop}%
\bibitem [{\citenamefont {Paley}\ and\ \citenamefont
  {Baharani}(2010)}]{paley2010critical}%
  \BibitemOpen
  \bibfield  {author} {\bibinfo {author} {\bibfnamefont {D.~A.}\ \bibnamefont
  {Paley}}\ and\ \bibinfo {author} {\bibfnamefont {A.~K.}\ \bibnamefont
  {Baharani}},\ }in\ \href@noop {} {\emph {\bibinfo {booktitle} {American
  Control Conference (ACC), 2010}}}\ (\bibinfo {organization} {IEEE},\ \bibinfo
  {year} {2010})\ pp.\ \bibinfo {pages} {4628--4633}\BibitemShut {NoStop}%
\end{thebibliography}%

\end{document}